%% file: main.tex
\documentclass[twocolumn]{aastex63}

\usepackage[T5,T1]{fontenc}
\usepackage{savesym}
\usepackage{amsmath}

\newcommand{\gcn}[1]{\href{https://gcn.gsfc.nasa.gov/gcn3/#1.gcn3}{GCN {#1}}}
\newcommand{\atel}[1]{\href{http://www.astronomerstelegram.org/?read=#1}{ATel {#1}}}

\shorttitle{IceCube Fast Response}
\shortauthors{Abbasi et al.}

\begin{document}

\title{Follow-up of astrophysical transients in real time with the IceCube Neutrino Observatory}

\input{authors.tex}

\collaboration{1000}{The IceCube Collaboration}
\noaffiliation
\begin{abstract}
In multi-messenger astronomy, rapid investigation of interesting transients is imperative. As an observatory with a 4$\pi$ steradian field of view and $\sim$99\% uptime, the IceCube Neutrino Observatory is a unique facility to follow up  transients, and to provide valuable insight for other observatories and inform their observing decisions. Since 2016, IceCube has been using low-latency data to rapidly respond to interesting astrophysical events reported by the multi-messenger observational community. Here, we describe the pipeline used to perform these follow up analyses and provide a summary of the 58 analyses performed as of July 2020. We find no significant signal in the first 58 analyses performed. The pipeline has helped inform various electromagnetic observing strategies, and has constrained neutrino emission from potential hadronic cosmic accelerators.
\end{abstract}


\section{Introduction} \label{sec:intro}
Recent successes of multi-messenger astronomy are due in large part to advancements in low-latency astronomical pipelines. Evidence for the first high-energy cosmic neutrino source, TXS 0506+056 \citep{2018Sci...361.1378I, 2018Sci...361..147I}, as well as the discovery of the first electromagnetic (EM) signal from a compact binary merger, GW170817, were both enabled by contemporaneous observations with various messengers \citep{2017PhRvL.119p1101A, 2017ApJ...848L..14G, 2017ApJ...848L..15S}. Observations of this type are made possible by public channels such as the Gamma-ray burst Coordinates Network (GCN)\footnote{\url{https://gcn.gsfc.nasa.gov/}} and the Astronomer's Telegram (ATel)\footnote{\url{http://www.astronomerstelegram.org/}}, which allow observers to coordinate observing strategies worldwide and quickly respond to interesting astrophysical transients. 

Among the myriad of questions being investigated with real-time multi-messenger astronomy is the identification of cosmic neutrino sources. Generated from the decay of charged pions that were created from proton-proton or photohadronic interactions in the vicinity of astrophysical accelerators,  neutrinos serve as excellent messenger particles. Whereas cosmic rays are deflected on their journey to Earth and high-energy photons produced in both leptonic and hadronic processes are attenuated by the extragalactic background light (EBL), neutrinos are  neither deflected nor attenuated, and provide a smoking gun signature for hadronic acceleration. 

\begin{figure*}[t]
    \centering
    \includegraphics[width=0.96\textwidth]{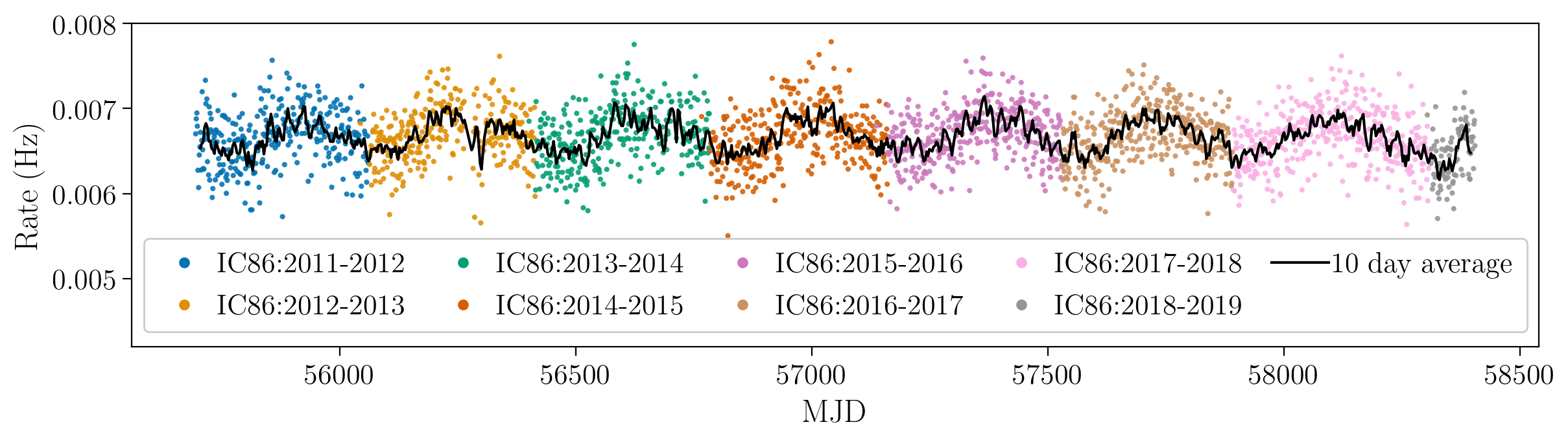}
    \caption{Rate of the GFU event selection over time. Different detector operation seasons are denoted by different colors, where ``IC86'' denotes the full 86 string detector configuration for IceCube. Each data point is the rate calculated from averaging 3 sequential 8-hour ``runs.'' As such, there is an expected Poissonian error for each data point on the order of 5\% from statistical fluctations only. In addition to this statistical fluctuation, the overall rate displays a clear annual modulation, whose peak to peak amplitude is approximately 4\% of the mean rate. To balance the effects of statistics and this annual modulation, the background rate is estimated using a running average with a 10 day width (black), described more fully in Section~\ref{sec:analysis}.}
    \label{fig:seasonal_variations}
\end{figure*}

However, the same small interaction probability which allows neutrinos to escape dense environments makes them notoriously difficult to detect. Additionally, cosmic rays interacting within Earth's atmosphere produce showers of particles, including atmospheric muons and neutrinos, which comprise a large background when searching for astrophysical neutrinos. Despite these challenges, a diffuse astrophysical neutrino flux has been detected \citep{2016ApJ...833....3A, 2013Sci...342E...1I, 2014PhRvL.113j1101A, 2020arXiv200109520I, Schneider:2019ayi, Stettner:2019tok}, and has been described with simple power laws from energies of about 10 TeV to 10 PeV. Although evidence for a first high-energy neutrino source has been presented, it is estimated that any neutrino flux from the object TXS 0506+056 could account for no more than 1\% of the total diffuse flux \citep{2020arXiv200109520I}.

In the search for astrophysical neutrino sources, correlation of signals in multiple channels is crucial, as the aforementioned atmospheric backgrounds are often overwhelming. In fact, in analyses searching for steady neutrino point sources, when looking at the entire sky with no a priori list of candidate objects, no neutrino source is significantly detected in 10 years of IceCube data \citep{2020PhRvL.124e1103A} as well as 11 years of ANTARES data \citep{2019arXiv190808248A}. It is not until neutrino data are correlated with lists of candidate neutrino emitters from EM observations that indications of neutrino signals from sources begin to manifest above the background expectation \citep{2020PhRvL.124e1103A}. However, attempts to correlate astrophysical neutrinos with known sources thus far have fallen short of explaining the diffuse flux, such as trying to corellate neutrinos with gamma-ray bursts (GRBs) \citep{2017ApJ...843..112A}, gamma-ray detected blazars \citep{Aartsen:2016lir}, fast radio bursts (FRBs) \citep{2020ApJ...890..111A, 2018ApJ...857..117A}, the Galactic plane \citep{2017ApJ...849...67A}, large-scale structure \citep{2019arXiv191111809I, Fang:2020rvq}, pulsar wind nebulae \citep{Aartsen:2020eof}, and the progenitors of gravitational waves \citep{2020ApJ...898L..10A, 2020arXiv200304022A, Albert:2018jnn, ANTARES:2017bia, Hussain:2019xzb, Keivani:2019smf}. Many of these searches have set strong constraints on source classes which were once believed to be dominant sources of astrophysical neutrinos.

However, the similarity in energy densities between the diffuse astrophysical neutrino flux and the extragalactic gamma-ray background observed by the Fermi telescope \citep{Ackermann:2014usa} may be suggestive of common origins \citep{Ahlers:2018fkn}. The lack of a clear correlation in previous catalog searches may indicate that while this correspondence may not be straightforward, subclasses of already investigated astrophysical sources could still be responsible for producing a significant fraction of the neutrino flux \citep{Halzen:2018iak}. Additionally, evidence for neutrino emission clustered in time during 2014-2015 from TXS 0506+056 \citep{2018Sci...361..147I} suggests that those extreme sources which may be neutrino emitters may also be variable in the time domain.

The identification of these extreme and variable sources is a problem well-posed for real-time observations. This was validated by the rigorous follow-up campaign of TXS 0506+056, as the neutrino alert on September 22, 2017 \citep{2017GCN.21916....1K} set off a multi-wavelength followup of over 20 telescopes across gamma-ray, X-ray, optical, and radio wavelengths. While the identification of potential neutrino sources based on pointing EM telescopes in the direction of neutrino alerts has already proven fruitful, one can also trigger followups using neutrino data to search for emission in the direction of EM objects while they are still in active states. This complementary approach provides another promising avenue for the identification of cosmic neutrino sources with real-time observations.

Here, we describe the fast-response analysis (hereafter FRA) pipeline established to rapidly search for neutrino emission from interesting astrophysical transients, using data from the IceCube Neutrino Observatory. This pipeline has been running since 2016, and a subset of the results were shared publicly via channels such as GCN and ATel to help inform EM observing strategies. We begin by providing a brief description of the data sample in Section~\ref{sec:GFU} and describe the analysis technique in Section~\ref{sec:analysis}. In Section~\ref{sec:targets}, we explain the types of sources which have been investigated with this pipeline, and then summarize all of our results as of July 2020 in Section~\ref{sec:results}.

\section{IceCube Data Sample} \label{sec:GFU}
\begin{figure*}
    \centering
    \includegraphics[width=0.96\textwidth]{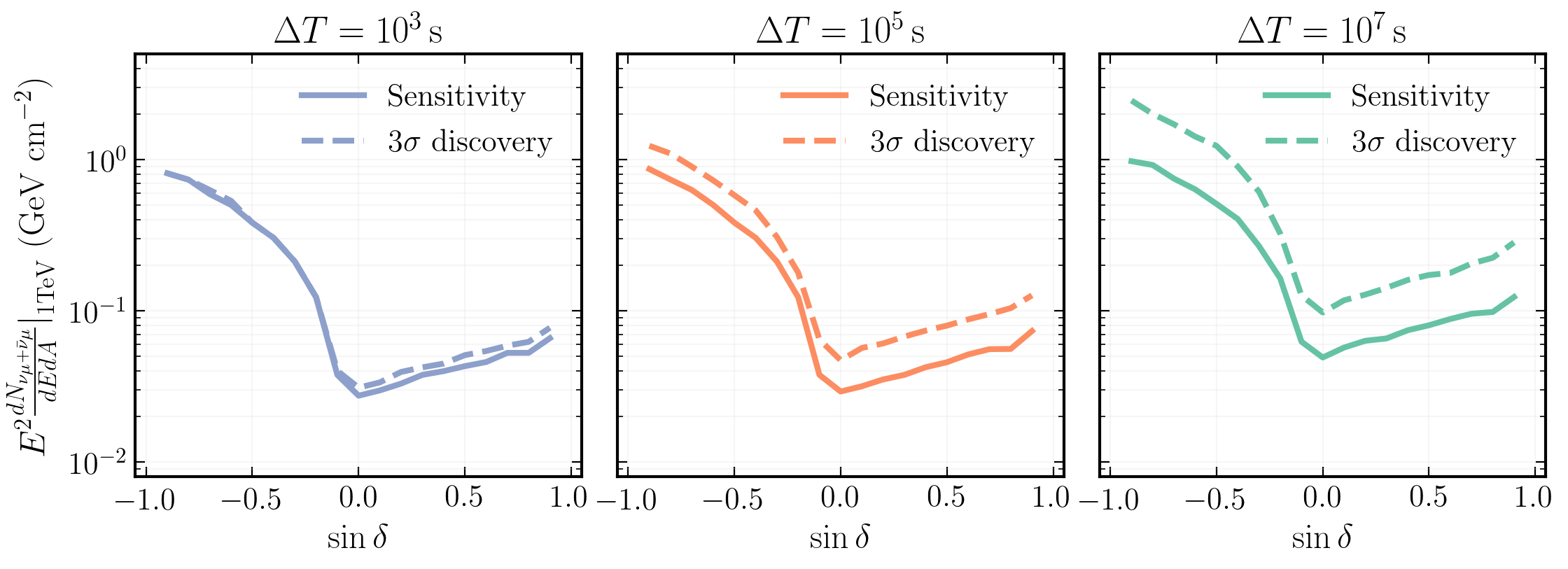}
    \caption{Analysis sensitivity as a function of declination ($\delta$) for characteristic analysis timescales of $10^3$ s (left), $10^5$ s (middle) and $10^7$ s (right), under the assumption of an $E^{-2}$ power-law spectrum. Sensitivity (solid line) is defined as the median 90\% CL upper limit that would be placed in the case of a non-detection, and discovery potential (dashed) as the flux required to yield a $3\sigma$ significant result, pre-trials, in 90\% of cases. The number of coincident neutrino candidate events increases as the time window for the analysis increases, which in turn increases the threshold for discovery. However, for time windows less than about one day, well-reconstructed individual coincident neutrino candidate  events are often capable of yielding analysis results that are significant at the $3\sigma$ level, pre-trials.}
    \label{fig:sensitivity_vs_declination}
\end{figure*}

The IceCube Neutrino Observatory is a cubic-kilometer neutrino detector instrumented at the geographic South Pole \citep{Achterberg:2006md, Aartsen:2016nxy}. The detector consists of 5160 digital optical modules (DOMs) distributed on 86 strings, and buried at depths between 1450 m and 2450 m. The DOMs consist of 10-inch photomultiplier tubes, onboard readout electronics, and a high-voltage board, all contained in a pressurized spherical glass container \citep{Abbasi:2008aa, Abbasi:2010vc}. In order to detect neutrinos, the DOMs can record Cherenkov radiation emitted by secondary particles produced by neutrino interactions in the surrounding ice or bedrock. Parameterization of the scattering and absorption of the glacial ice allows accurate energy and directional reconstruction of neutrino events~\citep{Aartsen:2013rt}.

IceCube's field of view covers the whole sky with $\sim$99\% uptime, though it is more sensitive to searches in the northern celestial hemisphere, where the Earth attenuates the majority of the atmospheric muon signal. Thus, the background at final selection level in the northern sky consists of atmospheric muon neutrinos from cosmic-ray air showers \citep{Haack:2017dxi}. In the southern sky the trigger rate is dominated by atmospheric muons from cosmic-ray air showers, and harsher cuts are placed to reduce this overwhelming background. 

Neutrino events in IceCube consist of two main morphologies: tracks and cascades. Tracks arise from charged-current $\nu_{\mu}$ interactions, wherein the produced muon creates a long ($\mathcal{O}$(km)) straight Cherenkov light pattern throughout the detector. Cascades, on the other hand, come from neutral current interactions of any flavor or charged-current $\nu_{e,\tau}$ interactions, and are characterized by spherical Cherenkov light patterns from particle showers. Whereas cascades have much better energy resolution, this analysis focuses on track-like events, as the long track topology not only provides an increased lever arm for better pointing resolution ($<1^{\circ}$ for $E_{\nu} > 10$ TeV), but also substantially increases the effective detection volume.

As this analysis runs in real time, it relies on the ability to have rapid access to data from the South Pole. Specifics of the infrastructure established to construct a real-time neutrino event stream are detailed in \cite{2017APh....92...30A}. This system has previously been used to rapidly send alerts to optical, X-ray, and gamma-ray telescopes \citep{Kintscher:2016uqh}, many of which have resulted in interesting EM observations \citep{Aartsen:2016qbu, Aartsen:2015trq, Abbasi:2011ja, Evans:2015qia}. Here, we focus on taking extreme transients from these EM observatories to prompt searches of our own data. The specifics of the event selection used here, which we refer to as the ``Gamma-ray Followup'' (GFU) dataset (because of its initial application in sending alerts to gamma-ray facilities), is described in full in \cite{Aartsen:2016qbu}. The angular resolution of the sample as a function of energy is displayed in \cite[Figure~4]{2017APh....92...30A}, and is very similar to other datasets that have been used for offline searches for neutrino sources \citep[Figure~1]{2020PhRvL.124e1103A}. At final level, the stream has an all sky rate that varies between approximately 6-7 mHz due to seasonal variations in the rate due to atmospheric backgrounds \citep{Desiati:2011hea, Tilav:2010hj, Grashorn:2009ey}. The variation of the sample's rate versus time is displayed in Fig.~\ref{fig:seasonal_variations}, and our treatment of this modulation is described in Section~\ref{sec:analysis}. This rate is dominated in the northern hemisphere by atmospheric neutrinos and in the southern hemisphere by atmospheric muons, but consists of $\mathcal{O}$(0.1\%) ($\mathcal{O}$(0.01\%)) neutrino candidate events (hereafter referred to as events) from astrophysical $\nu_{\mu}$ in the northern (southern) hemisphere.

\section{Analysis Method} \label{sec:analysis}

The FRA uses an unbinned maximum likelihood method that is also a feature in other IceCube searches for neutrino point sources \citep{Braun:2008bg, Braun:2009wp}, and preliminary forms of the analysis have been described in \citep{Meagher:2019edi, Meagher:2017htr}. For a sample with $N$ total neutrino candidate events in the analysis time window, we maximize the likelihood, $\mathcal{L}$, defined as

\begin{equation}
\begin{aligned}
    \label{eq:likelihood}
    \mathcal{L}(n_{s} | n_{b},&\left\{x_{i}\right\}) = \frac{\left(n_{s}+n_{b}\right)^{N} e^{-\left(n_{s}+n_{b}\right)}}{N !} \\ & \times \prod_{i=1}^{N}\left[ \frac{n_{s}}{n_{s}+n_{b}} \mathcal{S}\left(x_{i}\right)+\frac{n_{b}}{n_{s}+n_{b}} \mathcal{B}\left(x_{i}\right)\right],
\end{aligned}
\end{equation}
with respect to $n_s$, where $n_s$ and $n_b$ are the signal and expected background event counts, respectively. The term outside of the product in the likelihood is introduced to help distinguish potential signal events in the regime where the expected number of background events is small \citep{1990NIMPA.297..496B}. Maximizing this ``extended likelihood'' with respect to only $n_s$ has been a feature of previous IceCube analyses searching for short-timescale neutrino emission \citep{2017ApJ...843..112A, 2020ApJ...890..111A, 2020ApJ...898L..10A}. In Equation~\ref{eq:likelihood}, the index $i$ iterates over all neutrino event candidates, and $\mathcal{S}$ and $\mathcal{B}$ represent the signal and background probability density functions (PDFs) for events with observables $x_i$. The signal PDF, $\mathcal{S}$, is the product of both a spatial term, $\mathcal{S}_{\mathrm{space}}$, and an energy term, $\mathcal{S}_{\mathrm{energy}}$. The spatial term is modeled with a two dimensional Gaussian
\begin{equation}
    \mathcal{S}_{\mathrm{space}}=\frac{1}{2 \pi \sigma_{i}^{2}} e^{-\frac{\left|x_{s}-x_{i}\right|^{2}}{2 \sigma_{i}^{2}}} \; ,
\end{equation}
using the event's reconstruction uncertainty $\sigma_i$ for a source at location $x_s$. The energy term is used to distinguish background with a soft spectrum from signal with an assumed harder spectrum of $dN/dE \propto E^{-2}$. Thus, for each event, a PDF is evaluated using the event's energy proxy $E_i$ as well as its reconstructed declination, $\delta_i$, as the effective area of the sample has a strong dependence on declination.

Similarly, the background PDF, $\mathcal{B}$, is the product of a spatial term, $\mathcal{B}_{\mathrm{space}}$, and an energy term, $\mathcal{B}_{\mathrm{energy}}$. $\mathcal{B}_{\mathrm{space}}$ is estimated using experimental data, and depends only on the event's declination, as the probability in right ascension is treated as a uniform distribution $1/2\pi$. This yields
\begin{equation}
    \mathcal{B}_{\mathrm{space}} = \mathcal{P}_{\mathcal{B}}\left(\sin \delta_{i}\right) / 2 \pi \; ,
\end{equation}
where $\mathcal{P}_{\mathcal{B}}$ is the PDF of the sample as a function of declination, determined directly from experimental data. The background energy term is a two-dimensional PDF using the event's reconstructed declination and energy proxy, and is also determined directly from experimental data. 

The final test statistic, $\mathcal{T}$, is twice the logarithm of the ratio between the likelihood maximized with respect to $n_s$ (best-fit value $\hat{n}_s$) and that of the background-only likelihood ($n_s = 0$). This simplifies to

\begin{equation}
\label{eq:TS}
    \mathcal{T}=-2\hat{n}_{s}+2\sum_{i=1}^{N} \ln \left[\frac{\hat{n}_{s} \mathcal{S}\left(x_{i}\right)}{n_{b} \mathcal{B}\left(x_{i}\right)}+1\right] \; ,
\end{equation}
In order to determine $n_{b}$, we calculate the average rate in data in a time window 5 days in duration on either side of the time window being used for the analysis. For analyses being run in real time, there is often not 5 days of data available after the stop of the analysis time window, and for this we only use the 5 days of data leading up to the start of the analysis. The duration of 5 days was chosen such that it balances the uncertainty in rate between two competing effects: (1) the Poissonian uncertainty from the number of events detected and (2) the error from the fluctuating background rate due to seasonal variations, discussed in Section~\ref{sec:GFU}. We then keep $n_b$ fixed to this value, and only maximize the likelihood, $\mathcal{L}$, with respect to $n_s$.

The sensitivity of this analysis is dependent on the time window of the transient being investigated as well as its location on the sky, and we show the sensitivity for various characteristic time windows as well as different declinations in Figure~\ref{fig:sensitivity_vs_declination}. 

Sensitivities are defined assuming the flux is of the form

\begin{equation}
    \frac{dN_{\nu_{\mu} + \bar{\nu}_{\mu}}}{dE dA dt} = \phi_0 \times \Big(\frac{E}{E_0}\Big)^{-2}\;,
\end{equation}
and quoted in terms of the time-integrated flux, where $dN/dEdA = (dN/dEdAdt) \times \Delta T$, assuming constant emission. For short time windows, the analysis sensitivity is constant, as the expectation of having a coincident event from background is significantly less than one. In this regime, a single signal event is enough to yield a significant result in the analysis. Figure~\ref{fig:sensitivity_vs_declination} highlights the fact that the reduced background in the northern hemisphere significantly enhances the analysis sensitivity.

\begin{figure*}
    \centering
    \includegraphics[width=0.32\textwidth]{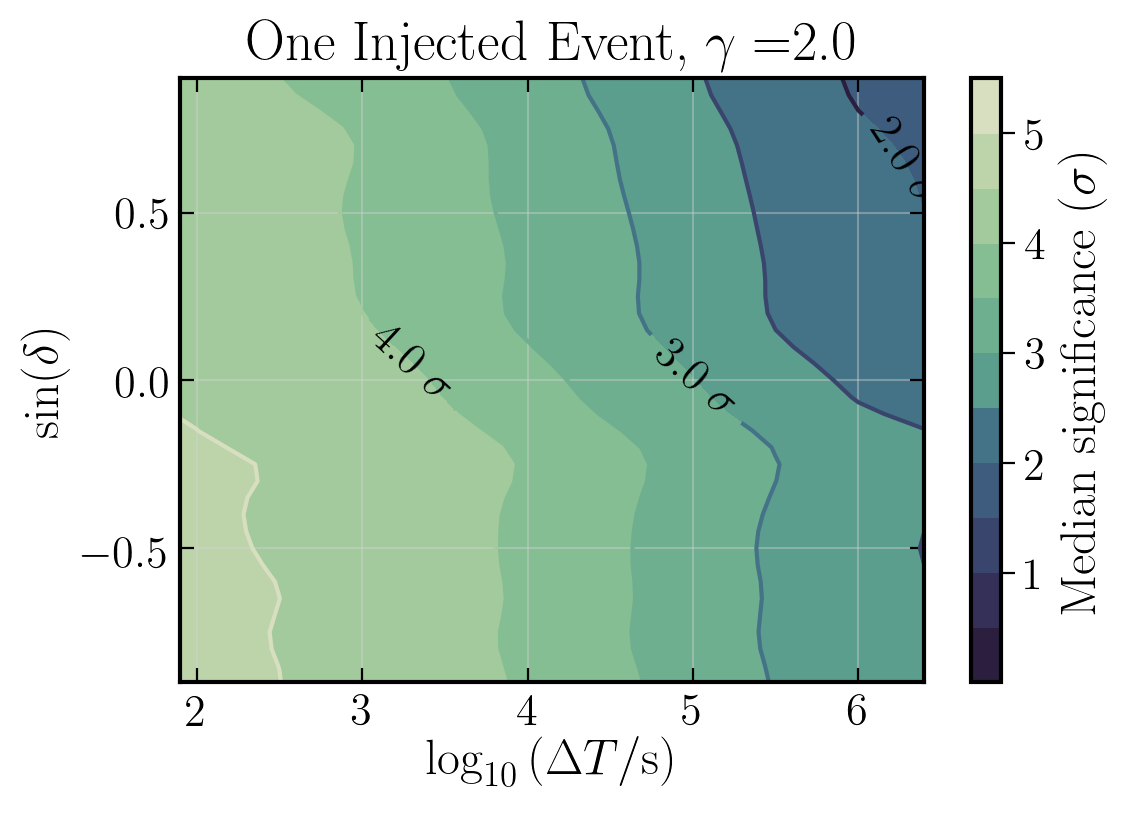}
    \includegraphics[width=0.32\textwidth]{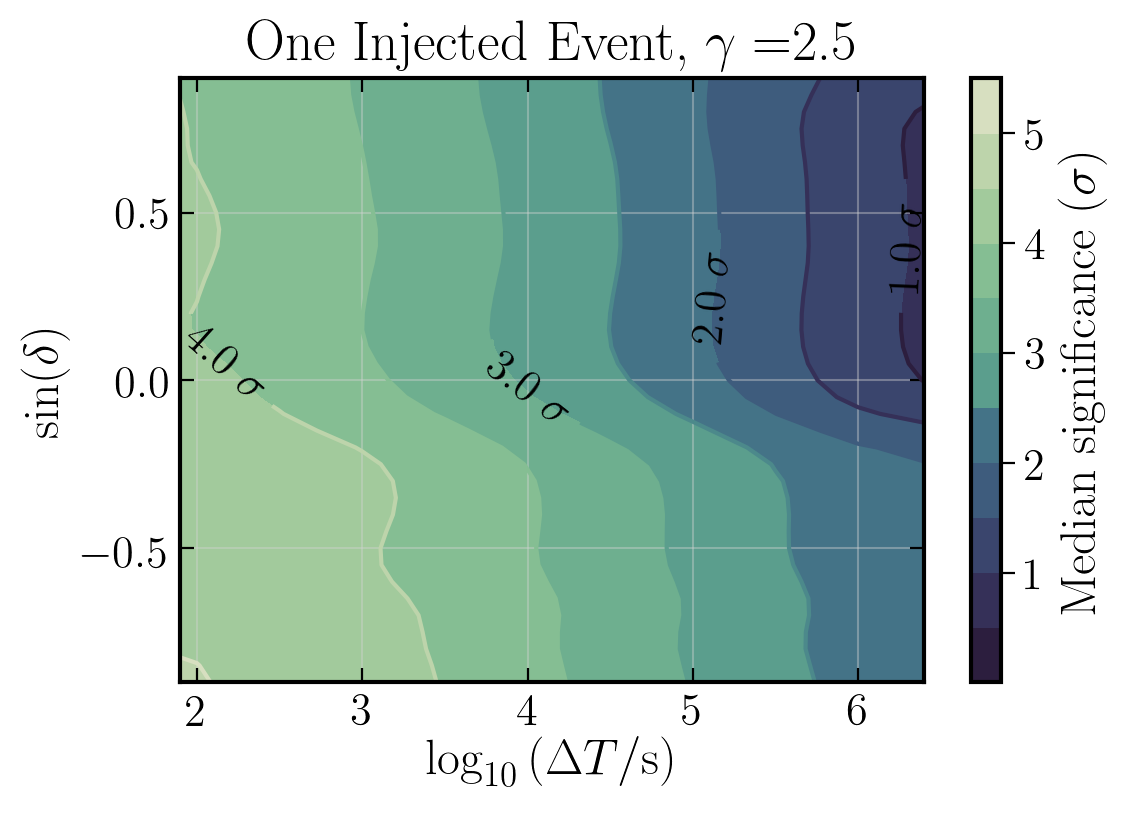}
    \includegraphics[width=0.32\textwidth]{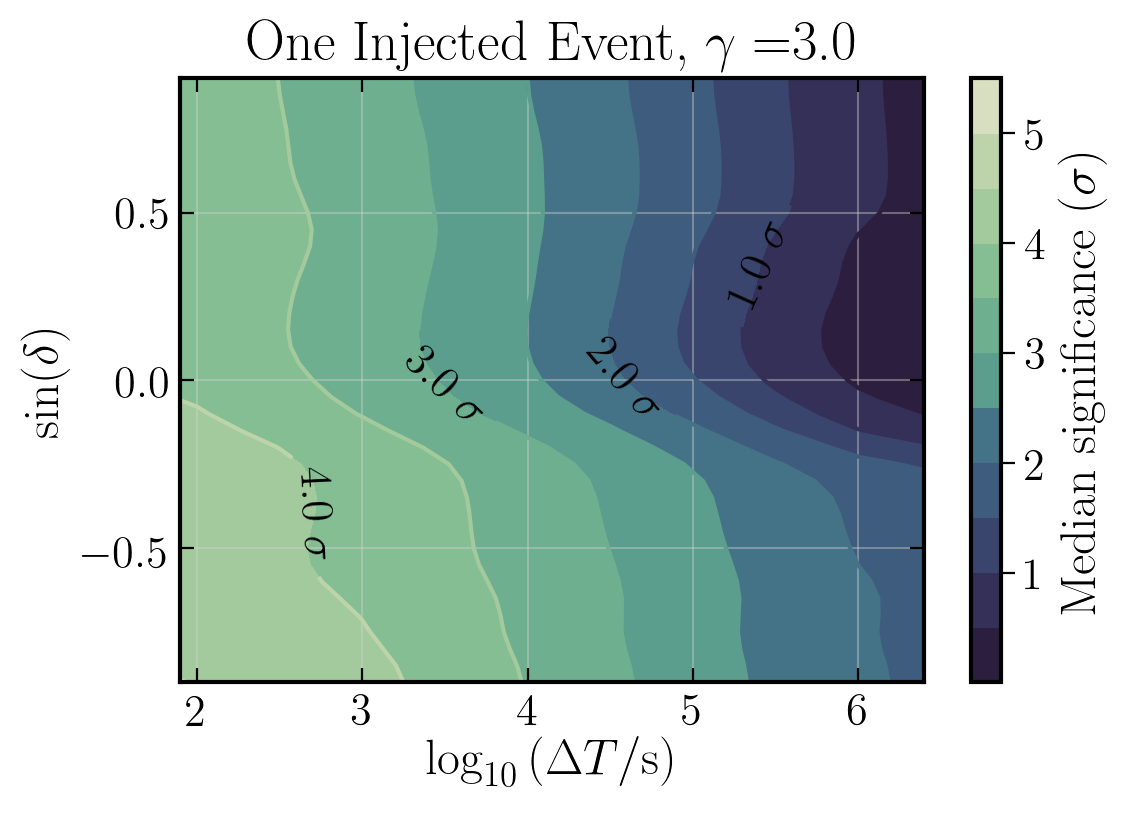}
    \caption{Statistical significance expected when detecting one signal neutrino candidate event. The colorscale represents the median pre-trial significance for analyses for a variety of timescales and declinations when there is one signal event, sampled according to an $E^{-\gamma}$ power-law spectrum for $\gamma=2.0$ (left), $\gamma=2.5$ (middle), and $\gamma=3.0$ (right), injected on top of scrambled background. Although the analysis is designed for incident $E^{-2}$ spectra, it remains sensitive to individual events from softer spectra. While a single event might result in a more significant result in the southern hemisphere than the northern hemisphere, the analysis has a much smaller effective area in the southern hemisphere, and is thus less sensitive in this hemisphere.}
    \label{fig:median_significance}
\end{figure*}

\begin{figure}
    \centering
    \includegraphics[width=0.44\textwidth]{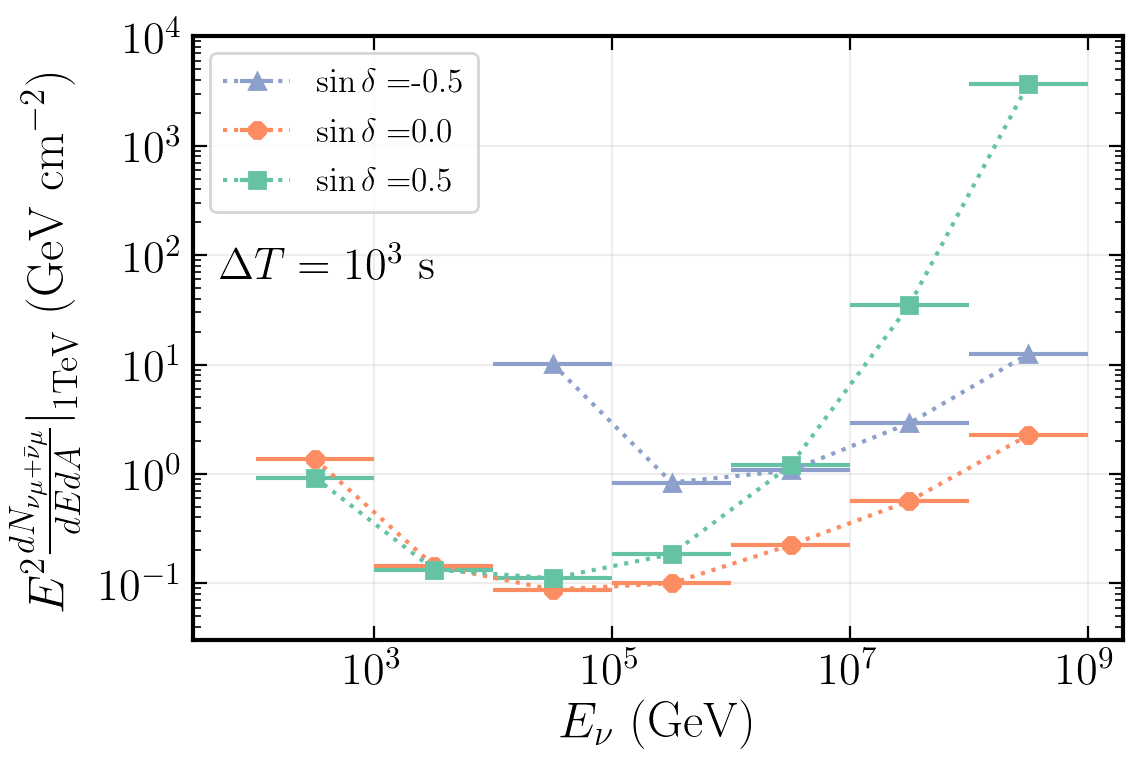}
    \caption{Differential sensitivity for an analysis with a $10^3$~s time window, for a source in the northern sky (green), at the horizon (orange), or in the southern sky (blue). The sensitivities are calculated separately in each decade in energy, assuming a differential muon neutrino flux $dN/dE\propto E^{-2}$ in that decade only. For events in the southern celestial hemisphere, the harsher cuts in the event selection make this analysis only sensitive at higher energies, whereas in the northern celestial hemisphere, the effect of Earth absorption is apparent in the highest energy bins.}
    \label{fig:differential_sensitivity}
\end{figure}

The advantage of this analysis in comparison with analyses which send alerts from IceCube is that it reduces the threshold needed for a detection. Analyses which send neutrino alerts either require high-energy neutrino candidates \citep{Blaufuss:2019fgv}, where the effective area is smaller than in the GFU sample so as to only select premier candidates, or  they require multiplets of lower-energy events in the GFU data \citep{Kintscher:2016uqh}. This analysis, however, is sensitive to individual events that do not need to be of the same quality or energy as IceCube alert events. The response of this analysis to individual neutrino candidate events is displayed in Figure~\ref{fig:median_significance} for different power-law spectra in terms of the median pre-trials significance over many realizations. The significance is calculated by comparing the observed $\mathcal{T}$ to those from pseudo-experiments in which the times of the events are scrambled \citep{Cassiday:1989kw, Alexandreas:1992ek}. The temporal scrambling preserves the detector acceptance as a function of declination, while altering the right ascension, and times are reassigned in such a way as to preserve the observed seasonal variations discussed in Section~\ref{sec:GFU}. For time windows larger than a few hours, the effective area and background rate in the GFU sample are independent of right ascension. For shorter time windows, the slightly asymmetric azimuthal geometry of the detector leads to an effective area and background rate that is up to 10\% higher for some right ascensions than others. Although this is not taken into account when calculating the signal and background PDFs, it does not introduce a bias in the calculation of the $p$-values which we report, as the temporal scrambling preserves local coordinates and thus maintains any azimuthal structure that is present in the sample.

Although the analysis is most sensitive to an incident $E^{-2}$ flux, it remains capable of yielding significant results if there is a source with a softer spectrum. Whereas other searches for point sources often fit the spectral index of any potential signal, e.g. \citep{2020PhRvL.124e1103A}, the index here is fixed, as we are looking for coincidences of individual events, from which it is not feasible to fit a spectrum.

As another way to highlight the analysis response to different spectral shapes, the differential sensitivity is provided in Figure~\ref{fig:differential_sensitivity}. The analysis is most sensitive at the celestial equator and northern sky for energies between $\mathcal{O}(10^3)$ GeV and $\mathcal{O}(10^5)$ GeV, whereas in the southern sky the harsher cuts increases this regime to be around $10^6$ GeV. For sources in the northern sky, Earth absorption becomes important at the highest energies.

\subsection{Sources with localization uncertainty}
\label{subsec:localization}
The analysis is also equipped to follow up sources where the uncertainty on the localization of the object is a significant fraction of the sky. This has application for searching for a variety of source classes, including but not limited to progenitors of gravitational waves, GRBs reported by the Fermi-GBM observatory, or poorly localized FRBs. In order to incorporate the localization uncertainty, the likelihood described in Equation~\ref{eq:likelihood} is maximized at every location on the sky, and the final test-statistic is defined as 

\begin{equation}
    \Lambda = \max_{\mathrm{\alpha, \delta}}\Bigg[ \mathcal{T}(\alpha, \delta) + 2\ln\Big(\frac{P_s(\alpha, \delta)}{P_s(\alpha_0, \delta_0)}\Big)\Bigg]\; ,
\end{equation}
where $\alpha, \delta$ are right ascension and declination, respectively. $P_s(\alpha, \delta)$ is the spatial PDF of the source being investigated, which consists of probabilities-per-pixel with pixels corresponding to locations on the sky generated according to the HEALPix scheme \citep{Gorski:2004by}. These PDFs are generally provided by the observatories which initially detect the transient of interest. $\alpha_0, \delta_0$ is the location on the sky corresponding to the maximum of this PDF, and $\mathcal{T}$ is the test-statistic defined in Equation~\ref{eq:TS}. This technique has also been used in dedicated analyses searching for counterparts to gravitational wave progenitors \citep{2020ApJ...898L..10A}, ANITA neutrino candidates \citep{2020ApJ...892...53A}, and ultra-high-energy cosmic rays \citep{Schumacher:2019qdx}.

\section{Follow-up Targets} \label{sec:targets}
In general, the FRA is run on extreme transients where there is potential for hadronic acceleration. Additionally, the analysis is used when it is believed that input from neutrino observations would be helpful in informing EM observing strategies. However, as the decision to perform the analysis is made on a case-by-case basis, it is difficult to define the exact circumstances that will result in an analysis. Potential targets predominately come from channels such as GCN or ATel, or are sometimes requested explicitly from EM observatories\footnote{Requests to perform the FRA can be sent to  \href{mailto:roc@icecube.wisc.edu}{roc@icecube.wisc.edu}}. In general, we favor sources that are detected with high-energy EM emission, and those sources which are in optimal locations for IceCube, namely, sources at or above the celestial equator. 

Once a potential target is identified, both the viability of the object being a neutrino emitter and the usefulness of input from a neutrino observatory for the EM community are evaluated. If it is decided to run the FRA, a time window, $\Delta T$, is selected that tries to encompass interesting periods of EM emission (for example, covering the entirety of a period of flaring activity reported in a GCN or ATel) while remaining in a regime where the analysis is most sensitive. After the analysis is complete, results are often shared via the channel where the emission that prompted the analysis was discussed.

As of July 2020, the FRA has been executed on a variety of astrophysical transients. While the analysis is designed to be applicable to generic objects, some classes of transients are followed up frequently (a complete list is provided in Appendix~\ref{app:results}). These classes include, though are not limited to: (1) extreme blazar flares, especially those detected in extremely high-energies, (2) bright GRBs, especially the few detected by imaging air cherenkov telescopes, (3) well-localized gravitational waves, (4) FRBs whose detections are released in real time, and (5) multi-messenger alert streams from the Astrophysical Multimessenger observatory Network\footnote{\url{https://gcn.gsfc.nasa.gov/amon.html}}. Since the pipeline's creation, some of these source classes have had dedicated real-time analyses, such as gravitational waves \citep{2020ApJ...898L..10A}. Dedicated real-time follow-ups of GRBs as well as the use of this pipeline to follow up neutrino candidate events sent by IceCube via AMON will be the subjects of future works.

\section{Results} \label{sec:results}
\begin{figure}
    \centering
    \includegraphics[width=0.44\textwidth]{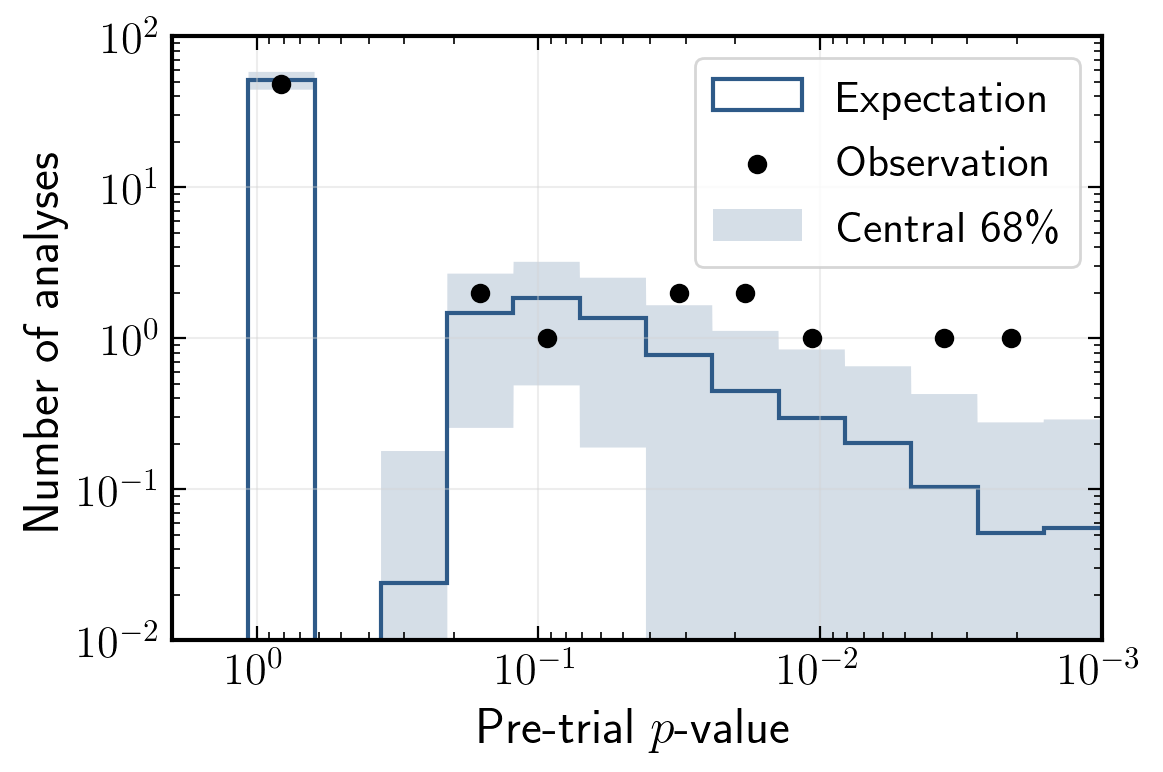}
    \caption{Distribution of $p$-values from all analyses. The $p$-values represent the outcome of each individual analysis, and do not include a trials correction for the ensemble of all analyses performed. As many of these analyses are looking for coincidences over short time windows, a large fraction of analyses have zero coincident events, yielding $\mathcal{T}=0$, and a $p$-value of exactly 1.0. We compare our distribution of $p$-values to those expected from many sets of ensembles of pseudo-experiments from scrambled background data for each of the 58 analyses performed.}
    \label{fig:pvalues}
\end{figure}

As of July 2020, the FRA has been used to follow up external observations 58 times. Although no analyses have resulted in significant results, we provide a complete list of results in Table~\ref{tab:results}. $p$-values are all quoted pre-trials, and upper limits are set assuming an $E^{-2}$ power law. For all analyses with $p<0.01$, we provide skymaps of the analysis in Appendix~\ref{app:skymaps}. A subset of these results were circulated via channels such as GCN or ATel, and links are provided where relevant. The distribution of all observed $p$-values is shown in Figure~\ref{fig:pvalues}. The background distribution of $p$-values is not expected to be perfectly uniform, as many analyses operated at short timescales, where there are zero observed coincident events. In this case, $\mathcal{T}=0$, and as this occurs for multiple pseudo-experiments, many pseudo-experiments yield the same value of $p=1.0$. As the hypotheses tested for the individual follow-up analyses are unique, we do not attempt to make any statement on the collection of results as a population, and instead we highlight some of the analyses individually in Section~\ref{sec:specific_details}.

For analyses with a $p$-value that is not 1.0, we find that the test-statistic is often dominated by one or two contributing neutrino candidate events. Although the analysis is capable of yielding significant results with one signal event from a hard astrophysical spectrum, none of our results are statistically significant as all of the coincident events had low reconstructed energies.

Some results that were shared via GCN or ATel prior to the writing of this work show slight differences in $p$-value as those presented here, as they were performed with a preliminary version of this analysis. The values provided in Table~\ref{tab:results} are all calculated with the analysis as described in Section~\ref{sec:analysis}. This version of the analysis has been stable since July 2020 and continues to operate in real time.

\subsection{Implications of specific analyses}
\label{sec:specific_details}
\begin{figure}
    \centering
    \includegraphics[width=0.44\textwidth]{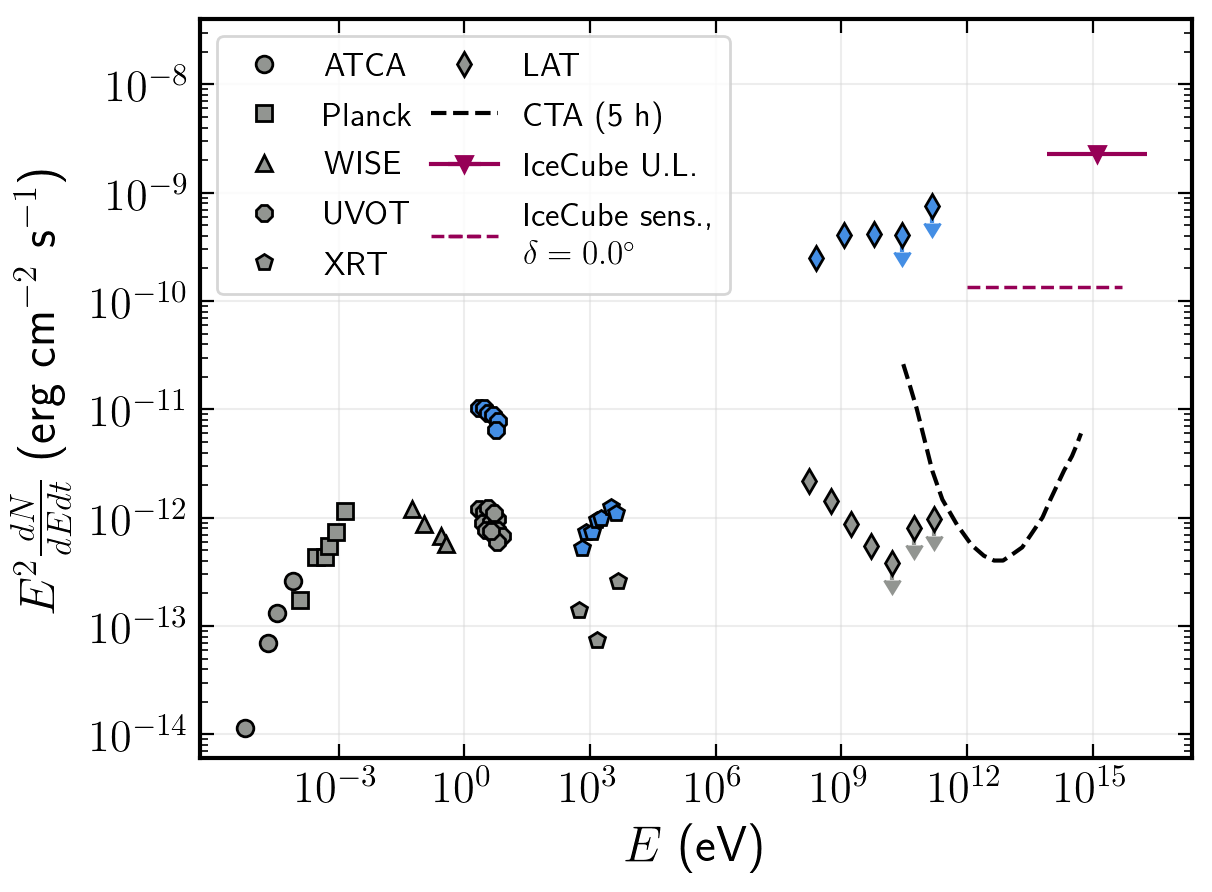}
    \caption{Spectral energy distribution of the flat spectrum radio quasar PKS 0346-27. All data points across the electromagnetic spectrum are taken from \cite{2019A&A...627A.140A}. Archival data are shown in gray, and data from May 16, 2018 are shown in blue. The limit placed by this analysis (solid magenta) uses a time window from May 11 - May 15, 2018, which covered the flaring activity on May 13, 2018 reported by the \textit{Fermi}-LAT (\atel{11644}). The May 16 time window for EM data points was chosen to have synchronous \textit{Swift} and \textit{Fermi}-LAT data. For comparison, we show the sensitivity (dashed magenta) this analysis would have for the same observation time window for a source at the horizon, where the sensitivity is optimal. The energies for both our upper limit and sensitivity span the central 90\% of expected energies assuming an $E^{-2}$ flux. The black dashed line shows the sensitivity for CTA south with 5 hours of observations, and is taken from \cite{2019scta.book.....C}.}
    \label{fig:pks_upper_limit}
\end{figure}

Below, we highlight some of the objects that were analyzed. Following each source name we include the declination of the object as well as the time window for the analyses performed, as these are the principle factors driving the analysis sensitivity:

\textbf{PKS 0346-27} ($\delta = -27.82^{\circ}$, $\Delta T = 4.2\times 10^5$ s): The most significant result comes from an analysis of the object PKS 0346-27, a flat spectrum radio quasar with redshift $z=0.991$. At the time of the analysis, the object was in a high state marked by a daily averaged gamma-ray flux approximately 150 times greater than its four-year average, and with at least one photon with $>30$ GeV energy detected by the \textit{Fermi}-LAT (\atel{11644}). Our analysis found one event coincident with the localization of PKS 0346-27, yielding a $p$-value of 0.0027, before correcting for the number of analyses performed. However, after trials correcting for the number of analyses performed, we note that this most significant analysis has a post-trials $p$-value of 0.145, which we find to be consistent with background. Our upper limits, compared to observations across the EM spectrum at the time of the flare \citep{2019A&A...627A.140A}, are displayed in Figure~\ref{fig:pks_upper_limit}. For this source, as it is located in the southern celestial hemisphere, we are only sensitive at the highest energies because of the strict cuts placed to reduce the harsh backgrounds in the southern sky. 

\textbf{AT 2018cow} ($\delta = +22.27^{\circ}$, $\Delta T = 3.0\times 10^5$ s): In recent years, time-domain optical surveys have revealed a growing class of rare and rapidly evolving extragalactic transients, or so-called ``Fast Blue Optical Transients (FBOTs), see e.g. \cite{Rest:2018qfk, Drout:2014dma, Arcavi:2015zie}. Among these objects is AT 2018cow, an object which prompted an extensive multi-wavelength follow-up campaign \citep{Margutti:2018rri}. Early in observations of the object, an FRA was run under the assumption that the object could be a Broad-Lined type Ic supernova, which has been considered as a potential source of astrophysical neutrinos \citep{2016PhRvD..93e3010T, 2016PhRvD..93h3003S, 2018ApJ...855...37D}. In this context, an analysis was performed with a 3-day time window, spanning the last optical non-detection to the first detection. Later observations of the object led to an array of possible classifications, including a tidal disruption event (TDE) or magnetar. In a separate analysis not part of the FRA program, the object was reanalyzed in the context of a potential TDE classification, implementing a time window from 30 days prior to peak to 100 days after \citep{Stein:2019ivm}. Although slight excesses were identified in both analyses, neither analysis was significant at even the $3\sigma$ level, pre-trials. As such, we claim no evidence of neutrino emission as neither analysis yielded statistically significant results. Magnetar based models of this object that also predict neutrino emission are noted to be significantly below the sensitivity of this analysis \citep{Fang:2018hjp}.

\begin{figure}
    \centering
    \includegraphics[width=0.44\textwidth]{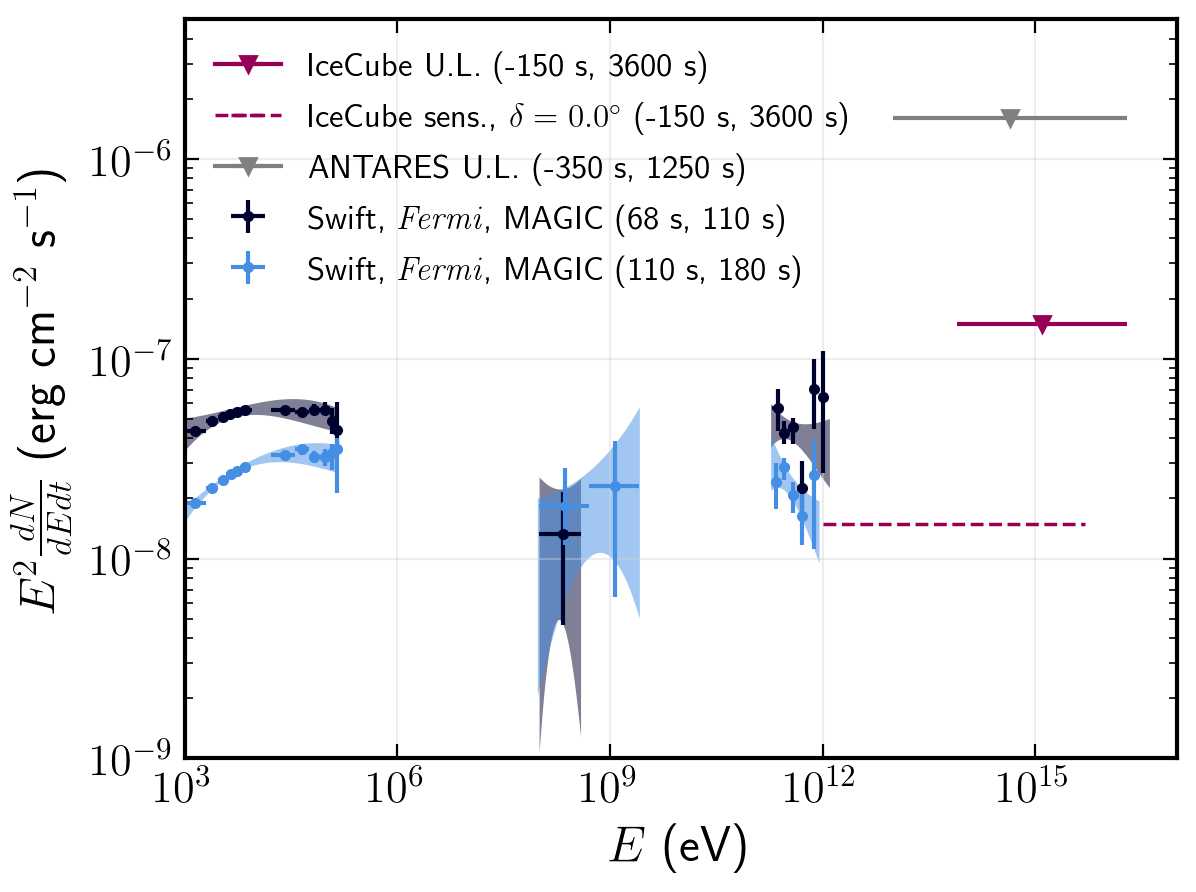}
    \caption{Multiwavelength and multimessenger spectra for GRB 190114C. Observation time windows are indicated in the legend. Neutrino upper limits are shown assuming an $E^{-2}$ flux, and span the central 90\% of the expected energies of neutrino events for this spectral assumption. The ANTARES limit is taken from \cite{Molla:2020neutrino}. Data points across the electromagnetic spectrum are taken from \cite{Acciari:2019dbx} and are shown for two time intervals. The lowest energy band  represents the 90\% confidence contours from a joint fit of Swift-BAT and Swift-XRT data, and the GeV and TeV bands are the 1$\sigma$ contour regions from the best-fit power-law functions from \textit{Fermi}-LAT and MAGIC, respectively. For comparison, we show the sensitivity (dashed magenta) this analysis would have for the same observation time window for a source at the horizon, where the sensitivity is optimal.}
    \label{fig:grb_190114c}
\end{figure}

\textbf{GRB 190114C} ($\delta = -26.94^{\circ}$, $\Delta T = 3.8\times 10^3$ s): This was the first GRB detected by an imaging air Cherenkov telescope that was announced in real time, with emission in the 0.2 - 1.0 TeV band detected by MAGIC \citep{Acciari:2019dxz}. Although the high energy peak in the broadband spectral energy distribution was later shown to be consistent with a synchrotron self-Compton interpretation \citep{Acciari:2019dbx}, GRBs have long been thought to be potential sources of astrophysical neutrinos \citep{1997PhRvL..78.2292W}. While no coincident events were observed, the southern declination of this GRB places it in a location of the sky where the event selection places stringent cuts to reduce the atmospheric muon background, see Figure~\ref{fig:differential_sensitivity}. As such, if there were neutrinos emitted at lower energies (less than $\mathcal{O}(10)$ TeV), the analysis would be much less sensitive than it would be for a similar source in the northern celestial hemisphere. The limits placed using this analysis are compared to the observations across the electromagnetic spectrum in Figure~\ref{fig:grb_190114c}. 

Given the redshift $z = 0.42$ and corresponding luminosity distance of approximately 2.3 Gpc of GRB 190114C \citep{Acciari:2019dbx}, we can also constrain the isotropic equivalent total radiated energy in muon neutrinos within our sensitive energy band, $E_{\nu, \mathrm{iso}}$. Using the upper limit presented in Table~\ref{tab:results}, we calculate 

\begin{equation}
    E_{\nu, \mathrm{iso}}=\frac{4 \pi D_{L}(z)^{2}}{1+z} \int_{E_{5\%}}^{E_{95\%}} \frac{dN^{90\%}_{\nu_{\mu} + \bar{\nu}_{\mu}}}{dE_{\nu} \,dA} E_{\nu} d E_{\nu}\; ,
\end{equation}
where $E_{5\%}$ and $E_{95\%}$ represent the bounds on the central 90\% of energies of detected events assuming an $E^{-2}$ spectrum, which for this declination we find to be around 100 TeV and 20 PeV, respectively. Accordingly, we constrain the total energy emitted in muon neutrinos within this energy range, assuming an $E^{-2}$ spectrum, to be less than $1.6\times 10^{54}$ erg (90\% CL). For comparison, the estimated isotropic energy emitted in photons was found to be around $3\times 10^{53}$ erg \citep{Acciari:2019dbx}. A similar calculation could be performed for any object that has a distance measurement as well as cataclysmic origins that we have investigated using the FRA. We have restricted our attention here to GRB190114C because of the extensive multi-wavelength observations of this object, and because it is one of the few GRBs detected at very high energies.

\textbf{SGR 1935+2154 / FRB 200428} ($\delta = +21.89^{\circ}$, $\Delta T = 8.6\times 10^4$ s): In April 2020, the CHIME/FRB instrument detected a millisecond timescale radio pulse coincident with a period of extraordinarily intense X-ray burst activity from a known Galactic magnetar, SGR 1935+2154 \citep{2020arXiv200510324T}, which was also detected by STARE2 \citep{2020arXiv200510828B}. Further analysis of the observables of this radio pulse, such as its duration and spectral luminosity, shows the signal to be indistinguishable from the expectation from an FRB, and this observation has supported the hypothesis that at least a fraction of the FRB population arise from magnetars \citep{2020arXiv200510828B}. Both magnetars, as well as FRBs, have been proposed as possible cosmic-ray accelerators \citep{2020arXiv200812318M, 2014ApJ...797...33L, 2018JApA...39...14G}, and as such, an analysis was performed searching for coincident neutrino events. The time window (2020-04-27 18:00:00 UTC to 2020-04-28 18:00:00 UTC) began approximately half an hour prior to the \textit{Swift}-BAT trigger (2020-04-27 18:26:20 UTC) and lasted 24 hours, covering all available data at the time of the analysis, and which encompassed the time of FRB 200428 (2020-04-28 14:34:24.45), which was approximately 20 hours after the start of this window. One coincident neutrino candidate event, arriving during the period of bursting X-ray activity (2020-04-27 19:23:30.93 UTC), but significantly before the FRB, was identified. This event had a relatively large uncertainty on its spatial reconstruction (2.67$^{\circ}$ at 90\% containment), and low reconstructed energy of $\sim 1$ TeV, which resulted in an analysis $p$-value of 0.02 (which is not corrected for the ensemble of all analyses performed), which we find to be not statistically significant. 

The results from this analysis, as well as other results that come from this pipeline, can be used to set limits on populations using extreme objects identified by EM observations, as we highlight below.

\begin{figure}
    \centering
    \includegraphics[width=0.46\textwidth]{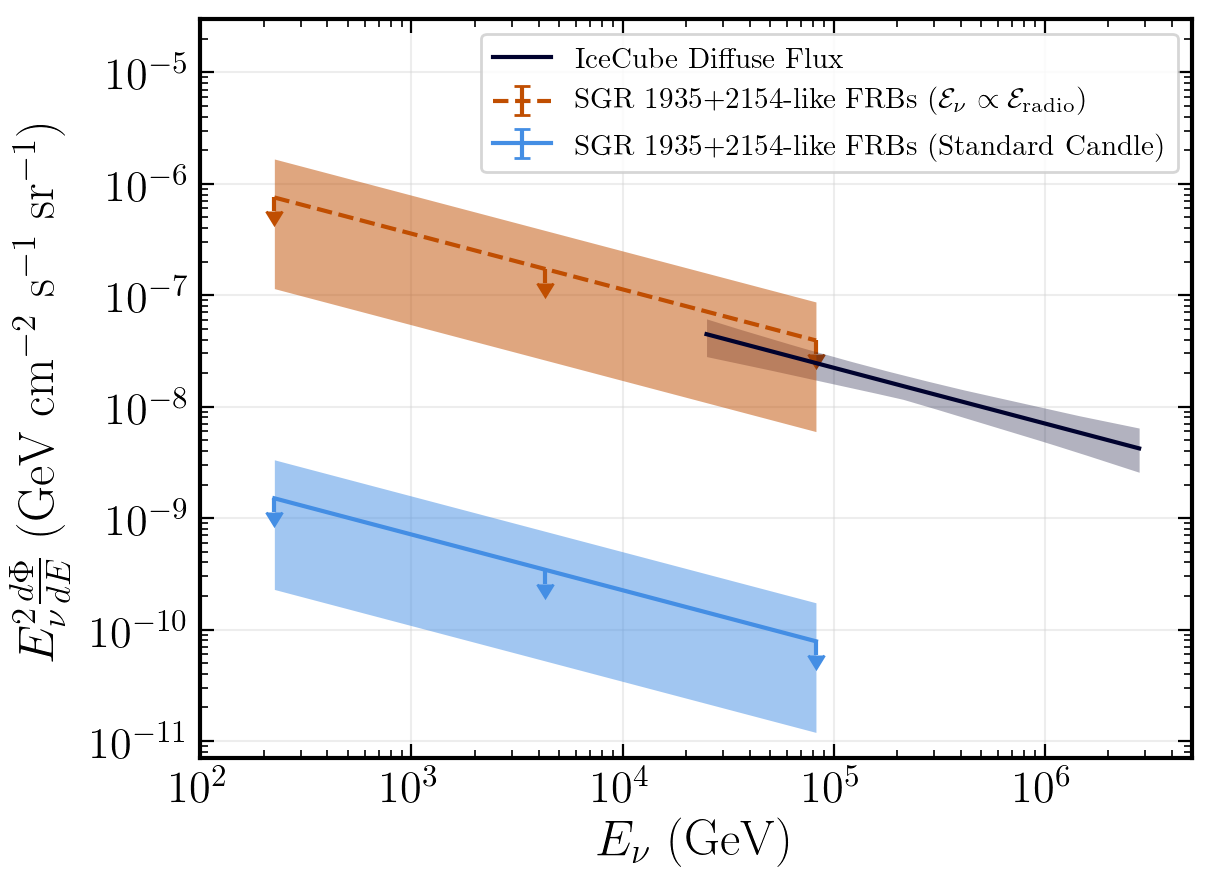}
    \caption{Upper limits on the contribution to the diffuse neutrino flux from a population of FRBs similar to SGR 1935+2154, for a variety of luminosity functions. The rate of such a population is taken from \cite{2020arXiv200510828B}, and the limit on the neutrino luminosity is derived from our analysis of FRB 200428. For a naive standard candle assumption (light blue), the strict upper limit from the Galactic burst limits the contribution of FRBs to be less than 1\% of the diffuse astrophysical neutrino flux. However, if the emitted energy in neutrinos were to scale linearly with the energy emitted at radio wavelengths (dashed orange), as described in the text, then FRBs are not ruled out as contributing significantly to the diffuse neutrino flux. The band on each of these limits represents the uncertainty on the reported volumetric rate of these transients.}
    \label{fig:frb_standard_candle}
\end{figure}

\cite{2020arXiv200510828B} showed that converting the detection of FRB 200428 to a volumetric rate of bursts results in an estimate of $7.23_{-6.13}^{+8.78} \times 10^{7} \mathrm{Gpc}^{-3} \mathrm{yr}^{-1}$ for this type of transient with energy greater than or equal to FRB 200428. We use this rate to set a constraint on the total contribution of FRBs from SGR 1935+2154-like bursts, assuming that for any neutrino flux, FRBs act as standard candles. An upper limit on this flux can be calculated using the technique outlined in \cite{nora_thesis}, namely, by integrating the rate of sources times their individual flux contributions over cosmic history

\begin{equation}
\label{eq:diff_FRB_cont}
    \frac{d\Phi}{dE} = \int_{0}^{\infty} R(z) \frac{dN}{dE} dz \; ,
\end{equation}
where $\frac{d\Phi}{dE}$ is the total diffuse differential flux from these bursts and $\frac{dN}{dE}$ is the differential flux from each source. $R(z)$ is the rate at which transients appear on Earth, given by
\begin{equation}
    R(z)=\rho(z) \times \frac{d V}{d z} \times \frac{1}{1+z} \; .
\end{equation}
For the volumetric rate density, $\rho(z)$, we use the rate discussed above and assume FRBs track star-formation activity, as is done in \cite{2020arXiv200510828B}. The other term in the integrand in Eq.~\ref{eq:diff_FRB_cont} is calculated as
\begin{equation}
    \frac{dN}{dE} = \frac{\mathcal{E}_{90\%}}{4 \pi D_{L}^{2}} \times(1+z)^{3-\gamma} E^{-\gamma} \; ,
\end{equation}
where  $\mathcal{E}_{90\%}$ is the upper limit on the time-integrated number of particles at 1 GeV of the burst released in neutrinos, assuming the emission follows a spectral shape consistent with the diffuse astrophysical neutrino spectrum as reported in \citep{2015ApJ...809...98A}. Although the likelihood used in the analysis assumes a spectral index of $\gamma=2.0$, the analysis is still sensitive when we calculate this limit by injecting a softer flux which has a spectral index of $\gamma=2.5$, as is shown in Figure~\ref{fig:median_significance}. To be conservative, we adopt a distance estimate for SGR 1935+2154 of 16 kpc, which was the maximal dispersion measure estimated distance reported in \cite{2020arXiv200510828B}. To calculate $\mathcal{E}_{90\%}$, we use the flux limit found using the FRA as well as the distance of SGR1935+2154. Our resulting limit, calculated with the public Flarestack code \citep{robert_stein_2020_4005800}, is displayed in Fig.~\ref{fig:frb_standard_candle}, which compares this upper limit to the total observed diffuse astrophysical neutrino flux. For SGR 1935+2154, this corresponds to a limit on the energy of the burst of $\sim 4\times 10^{43}$ erg emitted between energies of 200 GeV and 80 TeV for a neutrino flux of the form $dN/dE \propto E^{-2.5}$. We find that, under the assumption that FRBs that track star-formation activity and are standard candles in regards to their neutrino luminosities, that a population of FRBs with the aforementioned rate can contribute no more than 0.3\% of the diffuse neutrino flux. 

While the majority of the detected FRB population is extragalactic, a non-detection of a Galactic FRB implies an extremely small flux from extragalactic FRBs, under the assumption of standard candles. If, instead of assuming equal neutrino luminosities, the neutrino contribution were to scale linearly with the emitted radio energy, this constraint would scale by the ratio of the mean FRB energy to that from FRB 200428. If one assumes that the volumetric rate of FRBs per unit isotropic energy scales according to a power-law distribution $dN/d\mathcal{E}\propto \mathcal{E}^{-\gamma}$ with $\gamma=1.7$ and extends from the spectral energy of FRB 200428 out to a maximal spectral energy, $\mathcal{E}_{\mathrm{max}} \approx 2\times 10^{33} \mathrm{erg}\;\mathrm{Hz}^{-1}$ \citep{Lu:2019pdn}, then this ratio of spectral energies is on the order of $5\times 10^2$. Rescaling our upper limit on the total FRB contribution to the diffuse neutrino flux would then overshoot the total astrophysical neutrino flux, implying that if a population of SGR 1935+2154-like FRBs are not neutrino standard candles and instead have a positive correlation between neutrino and radio luminosities, then there is still room for them to significantly contribute to the diffuse neutrino flux. Even so, this limit highlights the fact that this pipeline can be used to constrain populations of potential neutrino sources by analyzing the most extreme objects identified in EM observations.

\section{Discussion \& Conclusion} \label{sec:discussion}
We presented a pipeline for rapidly investigating neutrino data in searches for extreme astrophysical transients. This analysis is well-suited to searching for individual coincident neutrinos with objects that were detected using other messengers. Since its start in 2016, this pipeline has proven useful in informing EM observers about possible neutrino emission, and have helped develop observing strategies. As of July 2020, no analyses have resulted in significant detections. Our limits have helped constrain various models of hadronic acceleration for a number of source classes that are thought to be cosmic-ray accelerators, including, but not limited to, superluminous transients such as AT2018cow and Galactic magnetars. The pipeline will continue to be operational. Beginning in 2018, this pipeline has circulated more of its results in real time via channels such as ATel or GCN, as is evident in Table~\ref{tab:results}. This has proven useful in aiding EM observing decisions, and these results have also been used by those creating lepto-hadronic emission models of certain transients of great interest to the observational community, such as AT2018cow \citep{Fang:2018hjp}.

With its $4\pi$ steradian field of view and $\sim$99\% uptime, IceCube is a unique observatory in that it is able to report on nearly every astrophysical transient. The ability to rapidly communicate a neutrino detected from an astrophysical transient enables the observational community to observe interesting objects as they are still in states of outburst, which could be pivotal in understanding the nature of astrophysical neutrinos.

\acknowledgments
{The IceCube Collaboration acknowledges the significant contributions to this manuscript from Alex Pizzuto and Justin Vandenbroucke. The authors gratefully acknowledge the support from the following agencies and institutions: USA {\textendash} U.S. National Science Foundation-Office of Polar Programs,
U.S. National Science Foundation-Physics Division,
Wisconsin Alumni Research Foundation,
Center for High Throughput Computing (CHTC) at the University of Wisconsin{\textendash}Madison,
Open Science Grid (OSG),
Extreme Science and Engineering Discovery Environment (XSEDE),
Frontera computing project at the Texas Advanced Computing Center,
U.S. Department of Energy-National Energy Research Scientific Computing Center,
Particle astrophysics research computing center at the University of Maryland,
Institute for Cyber-Enabled Research at Michigan State University,
and Astroparticle physics computational facility at Marquette University;
Belgium {\textendash} Funds for Scientific Research (FRS-FNRS and FWO),
FWO Odysseus and Big Science programmes,
and Belgian Federal Science Policy Office (Belspo);
Germany {\textendash} Bundesministerium f{\"u}r Bildung und Forschung (BMBF),
Deutsche Forschungsgemeinschaft (DFG),
Helmholtz Alliance for Astroparticle Physics (HAP),
Initiative and Networking Fund of the Helmholtz Association,
Deutsches Elektronen Synchrotron (DESY),
and High Performance Computing cluster of the RWTH Aachen;
Sweden {\textendash} Swedish Research Council,
Swedish Polar Research Secretariat,
Swedish National Infrastructure for Computing (SNIC),
and Knut and Alice Wallenberg Foundation;
Australia {\textendash} Australian Research Council;
Canada {\textendash} Natural Sciences and Engineering Research Council of Canada,
Calcul Qu{\'e}bec, Compute Ontario, Canada Foundation for Innovation, WestGrid, and Compute Canada;
Denmark {\textendash} Villum Fonden and Carlsberg Foundation;
New Zealand {\textendash} Marsden Fund;
Japan {\textendash} Japan Society for Promotion of Science (JSPS)
and Institute for Global Prominent Research (IGPR) of Chiba University;
Korea {\textendash} National Research Foundation of Korea (NRF);
Switzerland {\textendash} Swiss National Science Foundation (SNSF);
United Kingdom {\textendash} Department of Physics, University of Oxford.}
\software
{Flarestack~\citep{robert_stein_2020_4005800}, astropy~\citep{astropy}, numpy~\citep{numpy}, scipy~\citep{scipy} matplotlib~\citep{matplotlib}, pandas~\citep{pandas}
}

\bibliography{references}{}
\bibliographystyle{aasjournal}

\appendix
\section{List of Results}
\label{app:results}
The following table contains information on all of the analyses performed as of July 2020. References are provided to the GCN or ATel that prompted the analyses, though many of these objects were the topic of multiple GCN circulars or ATels. 

\include{results_table}

\onecolumngrid
\section{Skymaps}
\label{app:skymaps}
\begin{figure*}
    \centering
    \includegraphics[width=0.46\textwidth]{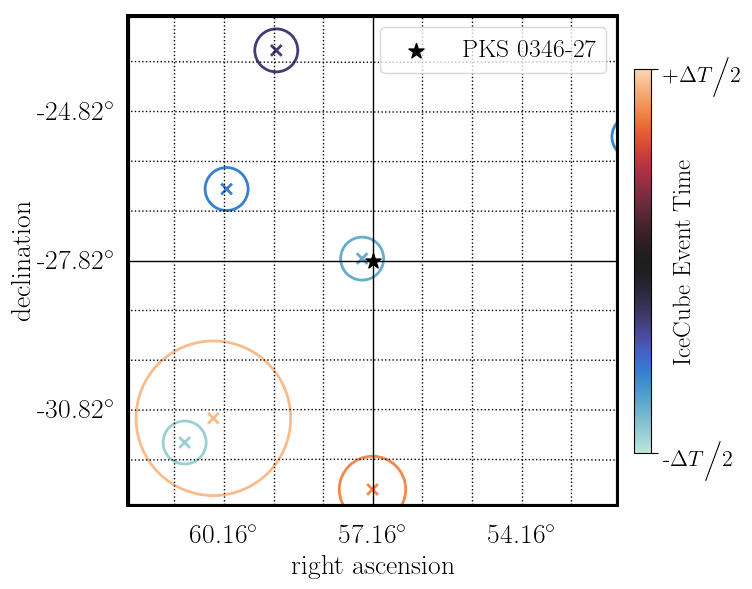}
    \includegraphics[width=0.46\textwidth]{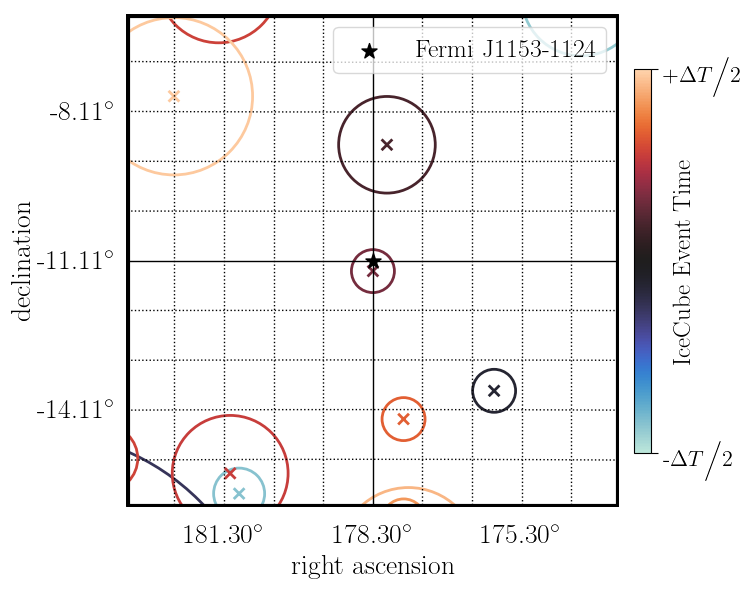}
    \caption{Skymaps for the two analyses which resulted in a $p$-value less than 0.01, pre-trials. Maps are centered on the location of the source being investigated. Each colored $\times$ represents a neutrino candidate event in the GFU sample, the color represents the arrival time of the event, and the size of the circle is each event's angular uncertainty (90\% containment). The analysis for PKS 0346-27 (left) lasted four days containing a period of increased gamma-ray emission detected by \textit{Fermi}-LAT and the analysis of a newly detected gamma-ray source Fermi J1153-1124 (right) was 2 days in duration.}
    \label{fig:skymaps}
\end{figure*}

In Figure~\ref{fig:skymaps}, we present skymaps of all analyses which resulted in a $p$-value less than 0.01, pre-trials, although we note that after trials corrections our most significant result is consistent with background, with a trials corrected $p$-value of 0.145. These analyses include (1) the follow-up of a bright GeV flare reported by the \textit{Fermi}-LAT from the blazar PKS 0346-27 and (2) Fermi J1153-1124, a source which, at the time, was a newly identified gamma-ray source.

\end{document}

%% file: authors.tex
\affiliation{III. Physikalisches Institut, RWTH Aachen University, D-52056 Aachen, Germany}
\affiliation{Department of Physics, University of Adelaide, Adelaide, 5005, Australia}
\affiliation{Dept. of Physics and Astronomy, University of Alaska Anchorage, 3211 Providence Dr., Anchorage, AK 99508, USA}
\affiliation{Dept. of Physics, University of Texas at Arlington, 502 Yates St., Science Hall Rm 108, Box 19059, Arlington, TX 76019, USA}
\affiliation{CTSPS, Clark-Atlanta University, Atlanta, GA 30314, USA}
\affiliation{School of Physics and Center for Relativistic Astrophysics, Georgia Institute of Technology, Atlanta, GA 30332, USA}
\affiliation{Dept. of Physics, Southern University, Baton Rouge, LA 70813, USA}
\affiliation{Dept. of Physics, University of California, Berkeley, CA 94720, USA}
\affiliation{Lawrence Berkeley National Laboratory, Berkeley, CA 94720, USA}
\affiliation{Institut f{\"u}r Physik, Humboldt-Universit{\"a}t zu Berlin, D-12489 Berlin, Germany}
\affiliation{Fakult{\"a}t f{\"u}r Physik {\&} Astronomie, Ruhr-Universit{\"a}t Bochum, D-44780 Bochum, Germany}
\affiliation{Universit{\'e} Libre de Bruxelles, Science Faculty CP230, B-1050 Brussels, Belgium}
\affiliation{Vrije Universiteit Brussel (VUB), Dienst ELEM, B-1050 Brussels, Belgium}
\affiliation{Department of Physics and Laboratory for Particle Physics and Cosmology, Harvard University, Cambridge, MA 02138, USA}
\affiliation{Dept. of Physics, Massachusetts Institute of Technology, Cambridge, MA 02139, USA}
\affiliation{Dept. of Physics and Institute for Global Prominent Research, Chiba University, Chiba 263-8522, Japan}
\affiliation{Department of Physics, Loyola University Chicago, Chicago, IL 60660, USA}
\affiliation{Dept. of Physics and Astronomy, University of Canterbury, Private Bag 4800, Christchurch, New Zealand}
\affiliation{Dept. of Physics, University of Maryland, College Park, MD 20742, USA}
\affiliation{Dept. of Astronomy, Ohio State University, Columbus, OH 43210, USA}
\affiliation{Dept. of Physics and Center for Cosmology and Astro-Particle Physics, Ohio State University, Columbus, OH 43210, USA}
\affiliation{Niels Bohr Institute, University of Copenhagen, DK-2100 Copenhagen, Denmark}
\affiliation{Dept. of Physics, TU Dortmund University, D-44221 Dortmund, Germany}
\affiliation{Dept. of Physics and Astronomy, Michigan State University, East Lansing, MI 48824, USA}
\affiliation{Dept. of Physics, University of Alberta, Edmonton, Alberta, Canada T6G 2E1}
\affiliation{Erlangen Centre for Astroparticle Physics, Friedrich-Alexander-Universit{\"a}t Erlangen-N{\"u}rnberg, D-91058 Erlangen, Germany}
\affiliation{Physik-department, Technische Universit{\"a}t M{\"u}nchen, D-85748 Garching, Germany}
\affiliation{D{\'e}partement de physique nucl{\'e}aire et corpusculaire, Universit{\'e} de Gen{\`e}ve, CH-1211 Gen{\`e}ve, Switzerland}
\affiliation{Dept. of Physics and Astronomy, University of Gent, B-9000 Gent, Belgium}
\affiliation{Dept. of Physics and Astronomy, University of California, Irvine, CA 92697, USA}
\affiliation{Karlsruhe Institute of Technology, Institute for Astroparticle Physics, D-76021 Karlsruhe, Germany }
\affiliation{Dept. of Physics and Astronomy, University of Kansas, Lawrence, KS 66045, USA}
\affiliation{SNOLAB, 1039 Regional Road 24, Creighton Mine 9, Lively, ON, Canada P3Y 1N2}
\affiliation{Department of Physics and Astronomy, UCLA, Los Angeles, CA 90095, USA}
\affiliation{Department of Physics, Mercer University, Macon, GA 31207-0001, USA}
\affiliation{Dept. of Astronomy, University of Wisconsin{\textendash}Madison, Madison, WI 53706, USA}
\affiliation{Dept. of Physics and Wisconsin IceCube Particle Astrophysics Center, University of Wisconsin{\textendash}Madison, Madison, WI 53706, USA}
\affiliation{Institute of Physics, University of Mainz, Staudinger Weg 7, D-55099 Mainz, Germany}
\affiliation{Department of Physics, Marquette University, Milwaukee, WI, 53201, USA}
\affiliation{Institut f{\"u}r Kernphysik, Westf{\"a}lische Wilhelms-Universit{\"a}t M{\"u}nster, D-48149 M{\"u}nster, Germany}
\affiliation{Bartol Research Institute and Dept. of Physics and Astronomy, University of Delaware, Newark, DE 19716, USA}
\affiliation{Dept. of Physics, Yale University, New Haven, CT 06520, USA}
\affiliation{Dept. of Physics, University of Oxford, Parks Road, Oxford OX1 3PU, UK}
\affiliation{Dept. of Physics, Drexel University, 3141 Chestnut Street, Philadelphia, PA 19104, USA}
\affiliation{Physics Department, South Dakota School of Mines and Technology, Rapid City, SD 57701, USA}
\affiliation{Dept. of Physics, University of Wisconsin, River Falls, WI 54022, USA}
\affiliation{Dept. of Physics and Astronomy, University of Rochester, Rochester, NY 14627, USA}
\affiliation{Oskar Klein Centre and Dept. of Physics, Stockholm University, SE-10691 Stockholm, Sweden}
\affiliation{Dept. of Physics and Astronomy, Stony Brook University, Stony Brook, NY 11794-3800, USA}
\affiliation{Dept. of Physics, Sungkyunkwan University, Suwon 16419, Korea}
\affiliation{Institute of Basic Science, Sungkyunkwan University, Suwon 16419, Korea}
\affiliation{Dept. of Physics and Astronomy, University of Alabama, Tuscaloosa, AL 35487, USA}
\affiliation{Dept. of Astronomy and Astrophysics, Pennsylvania State University, University Park, PA 16802, USA}
\affiliation{Dept. of Physics, Pennsylvania State University, University Park, PA 16802, USA}
\affiliation{Dept. of Physics and Astronomy, Uppsala University, Box 516, S-75120 Uppsala, Sweden}
\affiliation{Dept. of Physics, University of Wuppertal, D-42119 Wuppertal, Germany}
\affiliation{DESY, D-15738 Zeuthen, Germany}

\author{R. Abbasi}
\affiliation{Department of Physics, Loyola University Chicago, Chicago, IL 60660, USA}
\author{M. Ackermann}
\affiliation{DESY, D-15738 Zeuthen, Germany}
\author{J. Adams}
\affiliation{Dept. of Physics and Astronomy, University of Canterbury, Private Bag 4800, Christchurch, New Zealand}
\author{J. A. Aguilar}
\affiliation{Universit{\'e} Libre de Bruxelles, Science Faculty CP230, B-1050 Brussels, Belgium}
\author{M. Ahlers}
\affiliation{Niels Bohr Institute, University of Copenhagen, DK-2100 Copenhagen, Denmark}
\author{M. Ahrens}
\affiliation{Oskar Klein Centre and Dept. of Physics, Stockholm University, SE-10691 Stockholm, Sweden}
\author{C. Alispach}
\affiliation{D{\'e}partement de physique nucl{\'e}aire et corpusculaire, Universit{\'e} de Gen{\`e}ve, CH-1211 Gen{\`e}ve, Switzerland}
\author{A. A. Alves Jr.}
\affiliation{Karlsruhe Institute of Technology, Institute for Astroparticle Physics, D-76021 Karlsruhe, Germany }
\author{N. M. Amin}
\affiliation{Bartol Research Institute and Dept. of Physics and Astronomy, University of Delaware, Newark, DE 19716, USA}
\author{R. An}
\affiliation{Department of Physics and Laboratory for Particle Physics and Cosmology, Harvard University, Cambridge, MA 02138, USA}
% ...
\date{\today}

%% file: results_table.tex
\begin{longrotatetable}
\begin{deluxetable}{lccllccccll}
\tablecaption{Results of all fast response analyses to date. $p$-values are not trials corrected for the number of analyses performed. Upper limits are placed at the 90\% CL under the assumption of an $E^{-2}$ flux, and constrain the energy-scaled time-integrated flux, $E^2 dN/dEdA$. The energy range column denotes the central 90\% of the energies we would expect for signal events from a source at the given declination under the assumption of an $E^{-2}$ flux. Locations on the sky are quoted for the J2000 epoch. Analyses with no listed reference were triggered by private communications. \label{tab:results}}
\tablewidth{700pt}
\tabletypesize{\scriptsize}
\tablehead{
\colhead{Source Name} & \colhead{R.A.} & 
\colhead{dec.} & \colhead{Start time} & 
\colhead{Duration} & \colhead{$\hat{n}_s$} & 
\colhead{$-\log_{10}(p)$} & \colhead{Upper limit} & 
\colhead{Energy Range} & \colhead{Reference} & \colhead{IceCube Response} \\ 
\colhead{} & \colhead{($^{\circ}$)} & \colhead{($^{\circ}$)} & \colhead{(UTC)} & 
\colhead{(s)} & \colhead{} & \colhead{} &
\colhead{(GeV cm$^{-2}$)} & \colhead{(GeV)} & \colhead{} & \colhead{}
} 
\startdata
 Cygnus X-3 &  308.11 &  +40.96 &  2017-04-03 00:00:00.000 &  $8.64\times 10^{4}$ &  0.00 &  0.00 &  $5.2\times 10^{-2}$ &  ($7\times 10^{2}$,$4\times 10^{5}$) &  \atel{10243} &  {\textemdash} \\
 GRB 170405A &  219.83 &  -25.24 &  2017-04-05 18:35:49.000 &  $1.20\times 10^{3}$ &  0.00 &  0.00 &  $3.2\times 10^{-1}$ &  ($7\times 10^{4}$,$2\times 10^{7}$) &  \gcn{20987} &  {\textemdash} \\
 AGL J0523+0646 &  80.86 &  +6.78 &  2017-04-15 11:50:00.000 &  $4.32\times 10^{5}$ &  0.00 &  0.00 &  $3.9\times 10^{-2}$ &  ($1\times 10^{3}$,$3\times 10^{6}$) &  \atel{10282} &  {\textemdash} \\
 AT 2017eaw &  308.68 &  +60.19 &  2017-05-10 12:00:00.000 &  $2.59\times 10^{5}$ &  0.23 &  0.89 &  $7.4\times 10^{-2}$ &  ($6\times 10^{2}$,$2\times 10^{5}$) &  \atel{10372} &  {\textemdash} \\
 Fermi J1544-0649 &  236.08 &  -6.82 &  2017-05-15 00:00:00.000 &  $2.74\times 10^{5}$ &  0.00 &  0.00 &  $5.1\times 10^{-2}$ &  ($2\times 10^{3}$,$9\times 10^{6}$) &  \atel{10482} &  {\textemdash} \\
 Fermi J1544-0649 &  236.08 &  -6.82 &  2017-05-18 04:04:40.000 &  $9.36\times 10^{5}$ &  0.00 &  0.00 &  $5.6\times 10^{-2}$ &  ($2\times 10^{3}$,$9\times 10^{6}$) &  \atel{10482} &  {\textemdash} \\
 AXP 4U 0142+61 &  26.59 &  +61.75 &  2017-07-13 22:54:33.000 &  $7.20\times 10^{3}$ &  0.00 &  0.00 &  $5.9\times 10^{-2}$ &  ($5\times 10^{2}$,$2\times 10^{5}$) &  \gcn{21342} &  {\textemdash} \\
 GRB 170714A &  34.35 &  +1.99 &  2017-07-14 11:25:32.000 &  $4.36\times 10^{4}$ &  0.00 &  0.00 &  $3.0\times 10^{-2}$ &  ($1\times 10^{3}$,$5\times 10^{6}$) &  \gcn{21345} &  {\textemdash} \\
 AT 2017fro &  259.98 &  +41.68 &  2017-07-22 00:00:00.000 &  $1.21\times 10^{6}$ &  0.00 &  0.00 &  $6.1\times 10^{-2}$ &  ($7\times 10^{2}$,$3\times 10^{5}$) &  \atel{10652} &  {\textemdash} \\
 AGL J1412-0522 &  213.00 &  -5.40 &  2017-08-05 03:00:00.000 &  $1.73\times 10^{5}$ &  0.00 &  0.00 &  $4.0\times 10^{-2}$ &  ($2\times 10^{3}$,$8\times 10^{6}$) &  \atel{10623} &  {\textemdash} \\
 G298048 SSS17a &  197.45 &  -23.38 &  2017-08-17 12:32:44.000 &  $1.00\times 10^{3}$ &  0.00 &  0.00 &  $3.1\times 10^{-1}$ &  ($7\times 10^{4}$,$2\times 10^{7}$) &  \gcn{21529} &  {\textemdash} \\
 G298048 SSS17a &  197.45 &  -23.38 &  2017-08-17 12:41:04.000 &  $1.21\times 10^{6}$ &  0.00 &  0.00 &  $3.2\times 10^{-1}$ &  ($7\times 10^{4}$,$2\times 10^{7}$) &  \gcn{21529} &  {\textemdash} \\
 TXS 0506+056 &  77.36 &  +5.69 &  2017-09-15 00:00:00.000 &  $1.21\times 10^{6}$ &  0.00 &  0.00 &  $3.9\times 10^{-2}$ &  ($1\times 10^{3}$,$3\times 10^{6}$) &  \atel{10791}\footnote{The analysis of TXS 0506+056 reported by Fermi-LAT in this ATel was prompted by IceCube-170922A, an IceCube event which sent as a public alert. This event was excluded from the FRA analysis here.} &  {\textemdash} \\
 PKS 0131-522 &  23.27 &  -52.00 &  2017-11-16 00:00:00.000 &  $1.73\times 10^{5}$ &  0.77 &  1.39 &  $1.1\times 10^0$ &  ($9\times 10^{4}$,$2\times 10^{7}$) &  \atel{10987} &  {\textemdash} \\
 GRB 171205A &  167.41 &  -12.59 &  2017-12-05 06:20:43.000 &  $7.20\times 10^{3}$ &  0.00 &  0.00 &  $1.4\times 10^{-1}$ &  ($2\times 10^{4}$,$2\times 10^{7}$) &  \gcn{22177} &  {\textemdash} \\
 Mrk 421 &  166.11 &  +38.21 &  2017-12-19 00:00:00.000 &  $1.73\times 10^{5}$ &  0.00 &  0.00 &  $5.5\times 10^{-2}$ &  ($7\times 10^{2}$,$4\times 10^{5}$) &  \atel{11077} &  {\textemdash} \\
 Mrk 421 &  166.11 &  +38.21 &  2018-01-12 00:00:00.000 &  $8.64\times 10^{5}$ &  0.00 &  0.00 &  $5.9\times 10^{-2}$ &  ($7\times 10^{2}$,$4\times 10^{5}$) &  {\textemdash} &  {\textemdash} \\
 HESS J0632+057 &  98.25 &  +5.80 &  2018-01-17 00:00:00.000 &  $6.05\times 10^{5}$ &  0.00 &  0.00 &  $3.3\times 10^{-2}$ &  ($1\times 10^{3}$,$3\times 10^{6}$) &  \atel{11223} &  {\textemdash} \\
 CXOU J16740.2-455216 &  251.79 &  -45.87 &  2018-02-05 18:27:11.000 &  $6.88\times 10^{4}$ &  0.00 &  0.00 &  $6.3\times 10^{-1}$ &  ($9\times 10^{4}$,$2\times 10^{7}$) &  \atel{11264} &  {\textemdash} \\
 Sgr A* &  266.42 &  -29.01 &  2018-02-17 00:30:00.000 &  $1.80\times 10^{3}$ &  0.00 &  0.00 &  $3.7\times 10^{-1}$ &  ($8\times 10^{4}$,$2\times 10^{7}$) &  \atel{11313} &  {\textemdash} \\
 TXS 0506+056 &  77.36 &  +5.69 &  2018-03-09 00:00:00.000 &  $5.53\times 10^{5}$ &  0.00 &  0.00 &  $3.6\times 10^{-2}$ &  ($1\times 10^{3}$,$3\times 10^{6}$) &  \atel{11419} &  {\textemdash} \\
 FSRQ 3C 279 &  194.05 &  -5.79 &  2018-04-15 00:00:00.000 &  $3.02\times 10^{5}$ &  0.00 &  0.00 &  $3.8\times 10^{-2}$ &  ($2\times 10^{3}$,$8\times 10^{6}$) &  \atel{11545} &  {\textemdash} \\
 PKS 0346-27 &  57.16 &  -27.82 &  2018-05-11 00:00:00.000 &  $4.18\times 10^{5}$ &  0.98 &  2.57 &  $5.9\times 10^{-1}$ &  ($8\times 10^{4}$,$2\times 10^{7}$) &  \atel{11644} &  {\textemdash} \\
 PKS 0903-57 &  136.22 &  -57.58 &  2018-05-12 00:00:00.000 &  $3.31\times 10^{5}$ &  0.00 &  0.00 &  $8.1\times 10^{-1}$ &  ($1\times 10^{5}$,$2\times 10^{7}$) &  \atel{11644} &  {\textemdash} \\
 AT 2018cow &  244.00 &  +22.27 &  2018-06-13 00:00:00.172 &  $2.97\times 10^{5}$ &  1.19 &  1.64 &  $5.9\times 10^{-2}$ &  ($8\times 10^{2}$,$8\times 10^{5}$) &  \atel{11727} &  \atel{11785} \\
 2FHL J1037.6+5710 &  159.41 &  +57.17 &  2018-06-29 21:59:00.000 &  $1.69\times 10^{5}$ &  0.62 &  1.08 &  $8.3\times 10^{-2}$ &  ($6\times 10^{2}$,$2\times 10^{5}$) &  \atel{11806} &  {\textemdash} \\
 NVSS J163547+362930 &  248.95 &  +36.49 &  2018-07-06 12:00:00.000 &  $3.46\times 10^{5}$ &  0.00 &  0.00 &  $9.6\times 10^{-2}$ &  ($7\times 10^{2}$,$4\times 10^{5}$) &  \atel{11847} &  {\textemdash} \\
 FRB 180725A &  6.22 &  +67.05 &  2018-07-25 05:59:43.115 &  $8.64\times 10^{4}$ &  0.00 &  0.00 &  $7.7\times 10^{-2}$ &  ($5\times 10^{2}$,$1\times 10^{5}$) &  \atel{11901} &  {\textemdash} \\
 GRB 180728A &  253.57 &  -54.03 &  2018-07-28 16:29:00.073 &  $7.20\times 10^{3}$ &  0.00 &  0.00 &  $7.3\times 10^{-1}$ &  ($9\times 10^{4}$,$2\times 10^{7}$) &  \gcn{23046} &  {\textemdash} \\
 IGR J17591-2342 &  269.79 &  -23.71 &  2018-08-10 12:00:00.000 &  $1.50\times 10^{6}$ &  0.00 &  0.00 &  $3.2\times 10^{-1}$ &  ($7\times 10^{4}$,$2\times 10^{7}$) &  \atel{12004} &  {\textemdash} \\
 FSRQ 4C +38.41 &  248.82 &  +38.41 &  2018-09-01 09:00:00.000 &  $2.59\times 10^{5}$ &  0.00 &  0.00 &  $5.2\times 10^{-2}$ &  ($7\times 10^{2}$,$4\times 10^{5}$) &  \atel{12005} &  {\textemdash} \\
 HAWC All Sky Flare Alert &  101.82 &  +37.61 &  2018-09-02 11:22:30.000 &  $1.95\times 10^{5}$ &  0.00 &  0.00 &  $5.1\times 10^{-2}$ &  ($7\times 10^{2}$,$4\times 10^{5}$) &  {\textemdash} &  {\textemdash} \\
 AT 2018gep &  250.95 &  +41.05 &  2018-09-08 04:00:00.000 &  $1.42\times 10^{6}$ &  1.61 &  1.46 &  $1.2\times 10^{-1}$ &  ($7\times 10^{2}$,$4\times 10^{5}$) &  \atel{12030} &  \atel{12062} \\
 GRB 180914A &  52.74 &  -5.26 &  2018-09-14 11:31:47.000 &  $7.20\times 10^{3}$ &  0.00 &  0.00 &  $3.3\times 10^{-2}$ &  ($2\times 10^{3}$,$8\times 10^{6}$) &  \gcn{23225} &  {\textemdash} \\
 GRB 180914B &  332.45 &  +24.88 &  2018-09-14 18:22:00.000 &  $4.80\times 10^{2}$ &  0.00 &  0.00 &  $3.8\times 10^{-2}$ &  ($8\times 10^{2}$,$7\times 10^{5}$) &  \gcn{23226} &  {\textemdash} \\
 Crab nebula &  83.63 &  +22.01 &  2018-09-30 00:00:00.000 &  $1.03\times 10^{6}$ &  0.00 &  0.00 &  $5.4\times 10^{-2}$ &  ($8\times 10^{2}$,$9\times 10^{5}$) &  \atel{12095} &  {\textemdash} \\
 SDSS J00289.81+200026.7 &  7.12 &  +20.00 &  2018-10-03 12:00:00.000 &  $3.02\times 10^{5}$ &  0.00 &  0.00 &  $4.0\times 10^{-2}$ &  ($8\times 10^{2}$,$1\times 10^{6}$) &  \atel{12084} &  {\textemdash} \\
 Fermi J1153-1124 &  178.30 &  -11.11 &  2018-11-10 00:00:00.000 &  $1.73\times 10^{5}$ &  0.97 &  2.40 &  $1.6\times 10^{-1}$ &  ($9\times 10^{3}$,$1\times 10^{7}$) &  \atel{12206} &  \atel{12210} \\
 TXS 0506+056 &  77.35 &  +5.70 &  2018-11-27 00:00:00.000 &  $6.05\times 10^{5}$ &  0.00 &  0.00 &  $3.7\times 10^{-2}$ &  ($1\times 10^{3}$,$3\times 10^{6}$) &  \atel{12260} &  \atel{12267} \\
 GRB 190114C &  54.51 &  -26.94 &  2019-01-14 20:54:33.000 &  $3.78\times 10^{3}$ &  0.00 &  0.00 &  $3.5\times 10^{-1}$ &  ($8\times 10^{4}$,$2\times 10^{7}$) &  \atel{12390} &  \atel{12395} \\
 Mrk 421 &  166.08 &  +38.19 &  2019-04-08 00:00:01.000 &  $1.43\times 10^{6}$ &  0.00 &  0.00 &  $6.4\times 10^{-2}$ &  ($7\times 10^{2}$,$4\times 10^{5}$) &  \atel{12680} &  {\textemdash} \\
 ANTARES-LAT coincidence &  46.18 &  -8.27 &  2019-05-12 01:26:13.000 &  $2.00\times 10^{3}$ &  0.00 &  0.00 &  $6.7\times 10^{-2}$ &  ($3\times 10^{3}$,$1\times 10^{7}$) &  {\textemdash} &  {\textemdash} \\
 FRB 190711 &  329.00 &  -80.38 &  2019-07-10 13:53:41.100 &  $8.64\times 10^{4}$ &  0.00 &  0.00 &  $1.1\times 10^0$ &  ($9\times 10^{4}$,$2\times 10^{7}$) &  \atel{12922} &  \atel{12928} \\
 FRB 190711 &  329.00 &  -80.38 &  2019-07-11 01:52:01.100 &  $2.00\times 10^{2}$ &  0.00 &  0.00 &  $1.0\times 10^0$ &  ($9\times 10^{4}$,$2\times 10^{7}$) &  \atel{12922} &  \atel{12928} \\
 FRB 190714 &  183.97 &  -13.00 &  2019-07-13 17:37:12.901 &  $8.64\times 10^{4}$ &  0.00 &  0.00 &  $1.5\times 10^{-1}$ &  ($2\times 10^{4}$,$2\times 10^{7}$) &  \atel{12940} &  \atel{12956} \\
 FRB 190714 &  183.97 &  -13.00 &  2019-07-14 05:35:32.901 &  $2.00\times 10^{2}$ &  0.00 &  0.00 &  $1.4\times 10^{-1}$ &  ($2\times 10^{4}$,$2\times 10^{7}$) &  \atel{12940} &  \atel{12956} \\
 HAWC Burst Alert &  78.39 &  +6.61 &  2019-08-06 07:20:48.000 &  $4.32\times 10^{4}$ &  0.00 &  0.00 &  $3.5\times 10^{-2}$ &  ($1\times 10^{3}$,$3\times 10^{6}$) &  \href{https://gcn.gsfc.nasa.gov/notices_amon_hawc/1008846_957.amon}{GCN Notice} &  \gcn{25291} \\
 S190814bv &  13.95 &  -27.08 &  2019-08-14 18:46:39.010 &  $1.22\times 10^{6}$ &  0.00 &  0.00 &  $3.8\times 10^{-1}$ &  ($8\times 10^{4}$,$2\times 10^{7}$) &  \gcn{25487} &  \gcn{25557} \\
 GRB 190829A &  44.54 &  -8.97 &  2019-08-29 19:55:43.000 &  $9.00\times 10^{1}$ &  0.00 &  0.00 &  $8.5\times 10^{-2}$ &  ($4\times 10^{3}$,$1\times 10^{7}$) &  \gcn{25552} &  {\textemdash} \\
 GRB 190829A &  44.54 &  -8.97 &  2019-08-29 19:55:53.000 &  $8.64\times 10^{4}$ &  0.00 &  0.00 &  $8.7\times 10^{-2}$ &  ($4\times 10^{3}$,$1\times 10^{7}$) &  \gcn{25552} &  {\textemdash} \\
 HAWC-190917A &  321.84 &  +30.97 &  2019-09-16 19:14:19.000 &  $4.32\times 10^{4}$ &  0.93 &  1.85 &  $6.0\times 10^{-2}$ &  ($7\times 10^{2}$,$5\times 10^{5}$) &  \gcn{25766} &  \gcn{25775} \\
 1ES 2344+51.4 &  356.77 &  +51.71 &  2019-10-01 00:00:01.000 &  $5.18\times 10^{5}$ &  0.00 &  0.00 &  $7.0\times 10^{-2}$ &  ($6\times 10^{2}$,$3\times 10^{5}$) &  \atel{13165} &  {\textemdash} \\
 PKS 2004-447 &  301.98 &  -44.58 &  2019-10-22 12:00:00.000 &  $5.18\times 10^{5}$ &  0.00 &  0.00 &  $6.0\times 10^{-1}$ &  ($9\times 10^{4}$,$2\times 10^{7}$) &  \atel{13229} &  \atel{13249} \\
 SBS 1150+497 &  178.35 &  +49.52 &  2019-10-30 00:00:01.000 &  $1.73\times 10^{5}$ &  0.07 &  0.79 &  $5.7\times 10^{-2}$ &  ($6\times 10^{2}$,$3\times 10^{5}$) &  \atel{13253} &  \atel{13266} \\
 ANTARES-LAT coincidence &  240.45 &  -52.96 &  2020-01-26 18:52:02.950 &  $1.73\times 10^{5}$ &  0.00 &  0.00 &  $7.5\times 10^{-1}$ &  ($9\times 10^{4}$,$2\times 10^{7}$) &  \gcn{26915} &  \gcn{26922} \\
 VER J0521+211 &  80.44 &  +21.21 &  2020-02-25 02:52:48.000 &  $8.64\times 10^{4}$ &  0.00 &  0.00 &  $4.0\times 10^{-2}$ &  ($8\times 10^{2}$,$9\times 10^{5}$) &  \atel{13522} &  \atel{13532} \\
 PKS 0903-57 &  136.22 &  -57.59 &  2020-03-24 12:00:00.000 &  $1.81\times 10^{6}$ &  0.00 &  0.00 &  $7.5\times 10^{-1}$ &  ($1\times 10^{5}$,$2\times 10^{7}$) &  \atel{13632} &  {\textemdash} \\
 SGR 1935+2154 &  293.74 &  +21.89 &  2020-04-27 18:00:00.000 &  $8.64\times 10^{4}$ &  0.83 &  1.62 &  $5.5\times 10^{-2}$ &  ($8\times 10^{2}$,$9\times 10^{5}$) &  \atel{13675} &  \atel{13689} \\
\enddata
\end{deluxetable}
\end{longrotatetable}